\documentclass[%
 longbibliography,
 reprint,onecolumn,12pt,
 amsmath,amssymb,
 aps,
]{revtex4-1}

\usepackage{epsfig}
\usepackage{graphicx}% Include figure files
\usepackage{dcolumn}% Align table columns on decimal point
\usepackage{bm}% bold math
\usepackage{color} %%%YS
\usepackage{ulem}
\begin{document}
\newcommand{\ba}{\begin{array}}
	\newcommand{\ea}{\end{array}}
\newcommand{\be}{\begin{equation}}
	\newcommand{\ee}{\end{equation}}
\newcommand{\bea}{\begin{eqnarray}}
	\newcommand{\eea}{\end{eqnarray}}
\newcommand{\bfig}{\begin{figure}}
	\newcommand{\efig}{\end{figure}}
\newcommand{\Bl}{\Bigl}
\newcommand{\Br}{\Bigr}
\newcommand{\br}{{\bf r}}
\newcommand{\bV}{{\bf V}}
\newcommand{\RE}{{\rm Re}\,}
\newcommand{\IM}{{\rm Im}\,}
\newcommand{\re}{{\rm e}}
\newcommand{\ri}{{\rm i}}
\newcommand{\tF}{\tilde{F}}
\newcommand{\tG}{\tilde{G}}
\newcommand{\tI}{\tilde{I}}
\newcommand{\dd}{{\rm d}}
\newcommand{\al}{\alpha}
\newcommand{\Gm}{\Gamma}
\newcommand{\Lb}{\Lambda}
\newcommand{\vp}{\varphi}
\newcommand{\om}{\omega}
\newcommand{\omh}{\hat{\omega}}
\newcommand{\Om}{\Omega}
\newcommand{\Pih}{\hat{\Pi}}
\newcommand{\din}{\displaystyle\int\limits}
\newcommand{\IN}{\displaystyle\int\limits_{0}^}
\newcommand{\II}{\displaystyle\int\limits^{\infty}_}
\newcommand{\IE}{\displaystyle\int\limits^{1}_}

\preprint{APS/123-QED}

\title{Scattering of surface shallow water waves \\ on a draining bathtub vortex \protect \\{\small (Published in Phys. Rev. Fluids, 2019, v. 4, n. 3, 034704.)}}
%\title{Effect of Potential Vorticity on Scattering by a Draining Bathtub Vortex}
% Force line breaks with \\
%\thanks{A footnote to the article title}%

\author{Semyon Churilov}
\affiliation{Institute of Solar-Terrestrial Physics of the
Siberian Branch of Russian Academy of Sciences, Irkutsk-33, PO Box
291, 664033, Russia.}
\author{Yury Stepanyants}
\email{Corresponding author: Yury.Stepanyants@usq.edu.au}
\affiliation{Department of Applied Mathematics, Nizhny Novgorod
State Technical University, Nizhny Novgorod, 603950, Russia and \\
School of Agricultural, Computational and Environmental Sciences,
University of Southern Queensland, QLD 4350, Australia.}%

\date{\today}% It is always \today, today,
             %  but any date may be explicitly specified
\begin{abstract}
\hspace*{5cm}\\

In the linear approximation we study long wave scattering on an axially symmetric flow in a shallow water basin with a drain in the center. Besides of academic interest, this problem is applicable to the interpretation of recent laboratory experiments with draining bathtub vortices, description of wave scattering in natural basins, and also can be considered as the hydrodynamic analogue of scalar wave scattering on a rotating black hole in general relativity. The analytic solutions are derived in the low-frequency limit to describe both pure potential perturbations (surface gravity waves) and perturbations with nonzero potential vorticity. For the moderate frequencies the solutions are obtained numerically and illustrated graphically. It is shown that there are two processes governing the dynamics of surface perturbations, the scattering of incident gravity water waves by a central vortex, and emission of gravity water waves stimulated by a potential vorticity. Some aspects of their synergetic actions are discussed.
\end{abstract}

\pacs{Valid PACS appear here}% PACS, the Physics and Astronomy
                             % Classification Scheme.
%\keywords{Suggested keywords}%Use showkeys class option if keyword
                              %display desired
\maketitle

\section{INTRODUCTION}
\label{sec:1}
% \hspace*{\parindent}

Concentrated vortices, i.~e., localized flows with closed streamlines, represent ubiquitously existing formations in fluid flows; as the example one can mention atmospheric vortices, water whirlpools, vortices in liquid helium, plasmas and even galaxies) \cite{Batchelor, Dolotin}. The interaction of various types of waves (acoustic, surface and internal, Rossby waves) with vortices is one of the most essential hydrodynamical processes which usually determines the wave propagation in laminar and turbulent vortex flows. The development of the theory of wave scattering at localized vortices is the problem of a great importance with many practical applications. In particular, acoustic scattering by vortices is vital for the diagnostics of vortex flows and control of turbulence. There are vast publications in this field which is impossible to list in this paper; we only refer to the book \cite{StepFabr} where a reader can find references to many other publications. In the majority of cases studied thus far, the problem of wave scattering at concentrated vortices was considered for the cases when there is no water discharge from the basin. However, the problem with water discharge is very topical and has a number of practical applications to industry \cite{StepYeoh}, laboratory experiments \cite{Torres17}, and modelling of astrophysical phenomena \cite{ArtBH, AnalogGrav}. The latter circumstance is of special interest as it is closely related to the intriguing fundamental problem of Hawking radiation from black holes.

In 1981 Unruh \cite{Unruh81} established the analogy between the Hawking radiation emitted from the horizon of black holes and classical wave fields generated by inhomogeneous currents in continuous media. This analogy occurs both in terms of the physical phenomena and in the basic equations used for the description of the phenomena. Since that time many papers were published to study such phenomena in acoustics, hydrodynamics, optics, physics of condensed matter, etc. (see, for example, Refs. \cite{ArtBH,AnalogGrav} and references therein). In the simplest plane configuration, which corresponds to a non-rotating black hole, a great success has been achieved both in theoretical explanation of Hawking radiation \cite{Coutant-14, Robertson-16, Coutant-16, Philbin, ChuErStep} and in the experimental modelling of this phenomenon \cite{NJP08, NJP10, Weinfurtner, Euve, Steinhauer}.

Nowadays the main interest of researchers in this field is focused on the modelling of processes in the vicinity of rotating
	black holes. The distinguishing feature of such objects is that the
	event horizon is surrounded by the {\it ergosphere} within which any observer
	is inevitably co-rotating with the black hole \cite{Chandra}. Such a space-time configuration leads to a number of interesting physical phenomena; in particular, to amplification of various types of waves scattered on a rotating black hole, provided that the wave frequency $\om$ satisfies the condition (see, for example, \cite{PressT, Star, StarCh}):
	\be
	\om < m\Omega,
	\label{Super}
	\ee
where $\Om$ is the angular frequency of the black hole and $m$ is the azimuthal wave number. As has been shown in Refs. \cite{StepFabr, KopLeon}, under this condition the energy of vortex oscillations is negative which can lead to radiative instability of a vortex in a compressible fluid \cite{BroadMoore, GolFabr} and superradiance phenomenon playing an important role in various fields of physics \cite{Brito15}.

%	Fig. 1
\begin{figure}[b!]
	\centerline{\includegraphics[width=16cm]{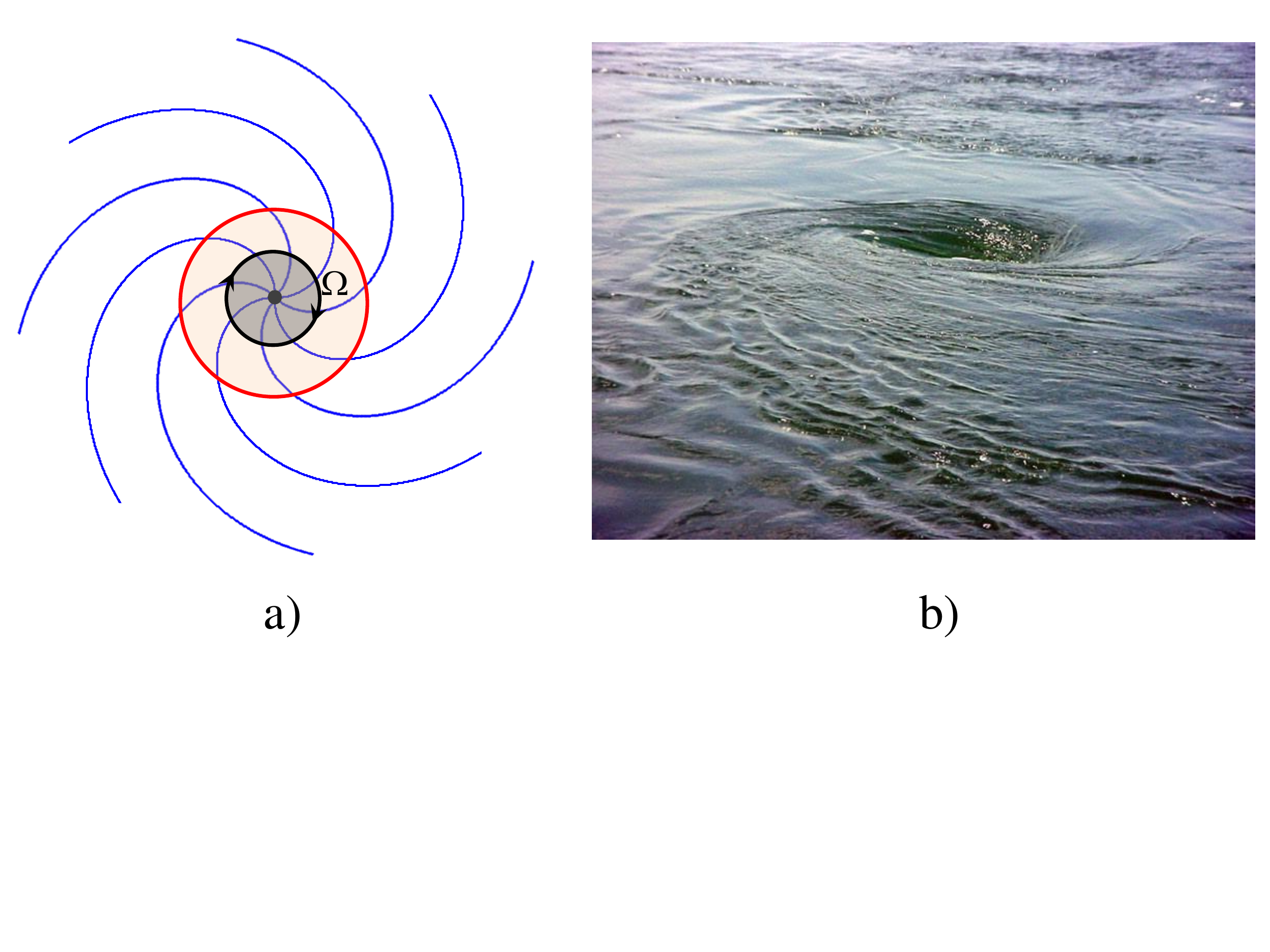}}%
\vspace*{-3.9cm}%
 	\caption{\protect\footnotesize Left frame (a) -- sketch of the wave pattern in the neighbourhood of a rotating black hole model: black spot in the centre shows a drainage orifice mimicking a black hole; black circle bounding the dark disk shows the event horizon; the red circle bounding the pink disk shows the ergosphere. Right frame (b) is a photo of a natural whirlpool ``Old Sow'' of about 80 m diameter that has existed for hundreds years off the coast of Maine, USA (image by Jim Lowe: {\color{blue} https://io9.gizmodo.com/a-250-foot-whirlpool-that-has-existed-for-hundreds-of-y-5501419?IR=T}).}
 	\label{f01}
 \end{figure}
 
  One of the simplest hydrodynamic models simulating both the event horizon and ergosphere is a draining bathtub (DBT) vortex \cite{SchUn, StepYeoh}. This is the axially symmetric flow of incompressible fluid in a shallow basin with a small drain hole in the centre (see Fig. \ref{f01}). The DBT flow has both the radial and azimuthal velocity components and can be considered as a two-dimensional flow everywhere, apart from the neighbourhood of the drain hole. Figure \ref{f01} illustrates a typical draining hole and a wave motion around it; the wave pattern impressed in the right frame (b) can be compared with the sketch shown in the left frame (a) which is similar to Fig.~2 of Ref.~\cite{VW05}.
 
 The DBT flow is potential because the circulation of its velocity around the centre is independent of the radius, and surface gravity waves in such flow are potential as well and described by the velocity potential $\phi$. The analogy with a scalar wave propagating in the neighbourhood of a rotating black hole is based on the fact that $\phi$ obeys the equation having the form of the d'Alembert equation in the {\it effective curved space-time} \cite{Dolan11, Dolan, Wein15}:
 \be
 \Box\phi\equiv\dfrac{1}{\sqrt{-g}}\dfrac{\partial}{\partial x^\mu}\left(
 \sqrt{-g}g^{\mu\nu}\dfrac{\partial\phi}{\partial x^\nu}\right)=0,
 \quad \mu,\,\nu= 0,\,1,\,2,
 \label{dAlamb}
 \ee
 where $g^{\mu\nu}$ is the (inverse) metric tensor and $g=1/\det(g^{\mu\nu})$ (should not be confused with the acceleration due to gravity used below).
 In recent years, scattering of surface waves by a DBT vortex was extensively studied
 and their amplification under condition (\ref{Super}) was calculated numerically (see, e.g., Refs.~\cite{Dolan, Wein15} and references therein) and confirmed in the laboratory experiments \cite{Torres17}.

 However, this effective metric, as well as those corresponding to other analogues based on the movements of continuous media, is not completely adequate to the Kerr metric \cite{Chandra} describing rotating black holes (see Ref.~\cite{VW05}, as well as a more comprehensive review \cite{Visser18} and numerous references therein). As mentioned in Ref. \cite{Visser18}, the reason is that the ``angular momentum in the physical spacetime to be mimicked corresponds to vorticity in the flow of the medium used in setting up analogue''. Meanwhile, vorticity of hydrodynamic flow leads to the linkage of various modes of its eigenoscillations resulting in that the d'Alembert equation in the effective curved space-time, describing wave propagation in continuous media, gains the right-hand side due to the mode coupling (for acoustic waves, this was demonstrated in detail in Ref.~\cite{Visser01}, see also Eqs.~(11) and (12) in Ref.~\cite{Visser18}).

An unperturbed DBT flow being the potential admits two kinds of disturbances, potential surface gravity waves and disturbances possessing potential vorticity (PV). The potential vorticity conserves and is simply transported by a flow as a passive impurity (see, for example, Ref. \cite{Dolzh} and Eq. (\ref{PV}) below). In this paper, we study scattering of both surface gravity waves and PV disturbances on a DBT flow to understand physically and describe quantitatively the result of their interplay.
 For low frequency waves, $\om \ll m\Om$, the problem can be solved analytically and we derive  corresponding solutions for surface waves propagating both with and without PV disturbances. Solutions in a wider range of frequencies, $\om \lesssim m\Om$, are found numerically. Our numerical results well agree with the analytical ones, when $\om \ll m\Om$.

The paper is organized as follows. In Section~\ref{sec:2} we derive the
 governing equations, describe the basic flow, formulate the boundary
 conditions for the perturbations, and establish the energy-flux
 conservation law for gravity waves. Axially symmetric ($m=0$)
 disturbances are studied in Section~\ref{sec:3} and it is found that in
 this case the potential vorticity has no effect on gravity waves.
 Section~\ref{sec:4} is devoted to the analytical consideration of
 low-frequency perturbations with $m > 0$ and it is shown that the
 disturbances carrying a potential vorticity stimulate emission of gravity
 waves; the emissivity is calculated. The numerical results are presented in Section~\ref{sec:5}, and Section~\ref{sec:6} is devoted to discussion of results obtained. In Appendix~\ref{ap:A} the
 derivation of conservation laws for gravity waves is given, and in
 Appendix~\ref{ap:B} some auxiliary calculations are presented.

\section{GOVERNING EQUATIONS, BASIC FLOW, AND BOUNDARY CONDITIONS}
\label{sec:2}
%\hspace*{\parindent}
%
In the shallow-water approximation, the basic set of hydrodynamic equations of perfect incompressible fluid with a free surface in the polar coordinates $(r,\vp)$ is (see, for example, \cite{Dolzh}):
\be
\left\{\ba{l}
\dfrac{\partial V_r}{\partial t}+V_r\,\dfrac{\partial V_r}{\partial r}+
\dfrac{V_\vp}{r}\,\dfrac{\partial V_r}{\partial\vp}-\dfrac{V_\vp^2}{r} =
-g\,\dfrac{\partial S}{\partial r},
\\ \\
\dfrac{\partial V_\vp}{\partial t}+V_r\,\dfrac{\partial V_\vp}{\partial r}+
\dfrac{V_\vp}{r}\,\dfrac{\partial V_\vp}{\partial\vp}+\dfrac{V_rV_\vp}{r} =
-\dfrac{g}{r}\,\dfrac{\partial S}{\partial\vp},
\\ \\
\dfrac{\partial H}{\partial t}+\dfrac{1}{r}\left[\dfrac{\partial}{\partial r}
(rHV_r)+\dfrac{\partial}{\partial\vp}(HV_\vp)
\right]=0.
\ea\right.
\label{BasEq}
\ee
where $g$ is the gravitational acceleration, $V_r$ and $V_\vp$ are the
velocity components of the fluid $\bm{V} = (V_r, V_\vp)$, $S(r,\vp,t)$ is the height of the free surface over a certain reference level, and $H(r,\vp,t)$ is the water depth. This set of equations conserves the potential vorticity (PV) \cite{Dolzh}:
\be
\Pi \equiv \dfrac{({\rm curl}\,\bm{V} )_z}{H}=\dfrac{1}{rH}\left[
\dfrac{\partial}{\partial r}\Bl(rV_\vp\Br)-\dfrac{\partial V_r}{\partial\vp}
\right], \qquad
\dfrac{\partial\Pi}{\partial t}+V_r\dfrac{\partial\Pi}{\partial r}+
\dfrac{V_\vp}{r}\dfrac{\partial\Pi}{\partial\vp}=0.
\label{PV}
\ee

It is easily seen that any steady axially symmetric flow $\bm{V}_0(\bm{r})$
with a sink at the origin is potential ($\Pi=0$) and can be described by a stream function $\Psi$, so that
\be
V_{r0}\equiv U(r)=\dfrac{\partial\Psi}{\partial r}, \quad V_{\vp 0}(r)
=\dfrac{1}{r}\,\dfrac{\partial\Psi}{\partial\vp}=\dfrac{C}{r},
\quad \mbox{where} \quad \Psi=\int\!U(r)\dd r + C\vp,
\label{BasFl}
\ee
and $C={\rm const}>0$ is the parameter, proportional to the velocity circulation ${\cal C} = 2\pi C$.

In a such flow the Bernoulli integral and mass-flux conservation law are fulfilled:
\be
\dfrac{1}{2}\left[U^2(r)+\dfrac{C^2}{r^2}\right]+gS_0(r)={\rm const},\ \ \ \
M = - rH_0(r)U(r)={\rm const}.
\label{Bernul}
\ee

Consider now a small perturbation of a potential flow with nonzero PV, in general:
\[
V_r=U(r)+u,\ \ \ \ rV_\vp=C+ \upsilon,\ \ \ \
H=H_0(r)+\eta,
% \label{Pert}
\]
where $|u(t, r, \vp)| \ll |U(r)|$, $| \upsilon(t, r, \vp)| \ll |rV_\vp|$, and $|\eta(t, r, \vp)| \ll |H_0(r)|$. We assume in this paper that the characteristic scale of basic flow and bottom profile variation in space is much greater than the wavelength of perturbation.

Linearizing Eqs.~(\ref{BasEq}), after simple manipulations we obtain two equations, one for the perturbation of angular momentum $\upsilon$, and another for the potential vorticity $\Pi$:
\bea
\label{PerEq}
\left[\hat{L} - U(r)\dfrac{H'_0(r)}{H_0(r)}\right]\hat{L} \upsilon &-&
\dfrac{g}{r}\left[\dfrac{\partial}{\partial r}\left(rH_0(r)
\dfrac{\partial \upsilon}{\partial r}\right)+\dfrac{H_0}{r}\dfrac{\partial^2 \upsilon}{\partial\vp^2}\right] = -\dfrac{g}{r}\dfrac{\partial}{\partial r}\Bl[r^2H_0^2(r)\,\Pi\Br], \\
\hat{L}\Pi &=& 0,
\label{Pi}
\eea
\[
\mbox{where  } \Pi = \dfrac{1}{rH_0(r)}\left(\dfrac{\partial \upsilon}{\partial r} -
\dfrac{\partial u}{\partial\vp}\right), \qquad
\hat{L} = \dfrac{\partial}{\partial t} + U(r)\dfrac{\partial}{\partial r} +
\dfrac{C}{r^2}\,\dfrac{\partial}{\partial\vp}.
\]
Note that Eq. (\ref{PerEq}) can be treated as the particular form of Lighthill's equation \cite{Goldstein}. Indeed, its left-hand side can be represented in the form of the d'Alambert Eq. (\ref{dAlamb}), and its right-hand side plays the role of the source term. When the right-hand side is zero (i.e., when $\Pi = 0$), Eq. (\ref{PerEq}) describes freely propagating gravity waves.

Following a traditional assumption used by many authors (see, for example, \cite{Dolan}), we also suppose that the water depth $H_0$ is independent of $r$. This dictates a certain dependence of $S_0(r)$ and corresponding bottom profile and also modifies the second Eq.~(\ref{Bernul}) in such a way, that $rU(r)=-F={\rm const}$, where $F$ is the drainage rate of the DBT vortex.

A particular solution of linearised Eq.~(\ref{PerEq}) can be considered in the form of a single harmonic of perturbation
$ \upsilon=f(r)\exp[\ri(m\vp-\omh t)]$, where $\omh >0$ is the frequency and $m$ is the azimuthal number; then we arrive to the equation:
$$
\left(\dfrac{F^2}{r^2}-gH_0\right)\dfrac{\dd^2 f}{\dd r^2}-
\left[gH_0-2\ri F\left(\omh-\dfrac{mC}{r^2}\right)+\dfrac{F^2}{r^2}
\right]\dfrac{1}{r}\,\dfrac{\dd f}{\dd r}
$$
\be
\label{Eq1}
{}-\left[\left(\omh-\dfrac{mC}{r^2}\right)^2-\dfrac{2\ri mCF}{r^4}-
\dfrac{m^2gH_0}{r^2}\right]f =
-\dfrac{g\,F^2}{r}\dfrac{\dd}{\dd r}\left(\dfrac{\Pi}{U^2(r)}\right).
\ee
This equation coincides with Eq.~(28) of the paper \cite{Wein15} in the particular case of $\Pi=0$ and $h\equiv H_0=$const).

As shown in Appendix~\ref{ap:A}, for potential perturbations with $\Pi=0$, the conserved radial energy flux is:
\be
rJ_{Er} = -2\omh\left(\omh - \dfrac{mC}{r^2}\right)F|f|^2 -
\ri\,\omh r\Bl[gH_0 - U^2(r)\Br]\left(f^*\dfrac{\dd f}{\dd r} -
f\dfrac{\dd f^*}{\dd r}\right) = {\rm const},
\label{ConsL}
\ee
where the asterisk denotes complex conjugation.

Let us introduce the dimensionless variables scaling velocity by $c_0=(gH_0)^{1/2}$ and spatial variable $r$ by the radius of the horizon $r_H = F/c_0$, then we have
\[
x=\dfrac{r}{r_H}\,, \quad \om=\dfrac{F\,\omh}{gH_0}=
\dfrac{r_H\,\omh}{c_0}\,, \quad
\Om=\dfrac{C}{F}=\dfrac{C}{c_0r_H}\,, \quad \Pih=\dfrac{F}{g}\,\Pi.
\]

Write down now the resultant equation for $f(r)$ in three equivalent forms: \\
- with the independent variable $x$:
\be
\hat{L}_x f\equiv
x^2(x^2-1)\dfrac{\dd^2 f}{\dd x^2}+\Bl[1+x^2-2\,\ri(\om x^2-m\Om)\Br]x
\dfrac{\dd f}{\dd x}+\Bl[(\om x^2-m\Om)^2 - m^2x^2-2\,\ri m\Om\Br]f=R_x;
\label{Eq-x}
\ee
- with the independent variable $y=x^2$:
\be
\hat{L}_y f\equiv
(y-1)\dfrac{\dd^2 f}{\dd y^2}+\left[1 - \ri\left(\om-\dfrac{m\Om}{y}
\right)\right] \dfrac{\dd f}{\dd y}+\dfrac{1}{4}\left[
\left(\om-\dfrac{m\Om}{y}\right)^2 - \dfrac{2\ri m\Om}{y^2}-
\dfrac{m^2}{y}\right]f = R_y;
\label{Eq-y}
\ee
- and with the independent variable $s=1/y$:
\be
\hat{L}_s f\equiv
s^3(1-s)\dfrac{\dd^2 f}{\dd s^2}-\Bl[2s-1 - \ri(\om-m\Om s)\Br]s^2\,
\dfrac{\dd f}{\dd s}+\dfrac{1}{4}\Bl[(\om-m\Om s)^2 - 2\ri m\Om s^2-
m^2s\Br]f = R_s,
\label{Eq-s}
\ee
where
\[
R_x = x^3\dfrac{\dd}{\dd x}\Bl(x^2\Pih\Br), \quad
R_y = \dfrac{1}{2}\,\dfrac{\dd}{\dd y}\Bl(y\Pih\Br), \quad
R_s = -\dfrac{s^2}{2}\,\dfrac{\dd}{\dd s}\left(\dfrac{\Pih}{s}\right).
\]

Equation~(\ref{Pi}) for PV, $\hat{L}\Pih=0$, is easily integrated:
\[
\Pih = \Pi_0x^{\ri m\Om}\re^{-\ri\om x^2/2} =
\Pi_0y^{\ri m\Om/2}\re^{-\ri\om y/2} =
\Pi_0s^{-\ri m\Om/2}\re^{-\ri\om/(2s)}, \quad \Pi_0={\rm const},
\]
so that
\be
\ba{l}
R_x=-\ri\Pi_0(\om x^2-m\Om+2\ri)x^{\ri m\Om+4}\re^{-\ri\om x^2/2},
\\ \\
R_y=-\dfrac{\ri\Pi_0}{4}\Bl(\om y-m\Om+2\ri\Br)y^{\ri m\Om/2}
\re^{-\ri\om y/2},
\\ \\
R_s=-\dfrac{\ri\Pi_0}{4}\Bl(\om-m\Om s+2\ri s\Br)s^{-\ri m\Om/2-1}
\re^{-\ri\om/(2s)}.
\ea
\label{Rxys}
\ee

The general solution of any of the equivalent equations (\ref{Eq-x}) -- (\ref{Eq-s}) is the sum of the general solution of the homogeneous equation and a particular solution of the non-homogeneous equation. Let us start with homogeneous equations. Near the horizon, which is their regular singular point, it is more convenient to use Eq.~(\ref{Eq-y}). The Frobenius expansions of its linear independent solutions have the following asymptotic forms up to $O[(y-1)^2]$ \cite{Ince}:
\begin{eqnarray}
f_a(y) &=& 1-\dfrac{(\om-m\Om)^2 - 2\ri m\Om - m^2}{4[1 - \ri(\om-m\Om)]}\,(y-1), \label{FrobIn1} \\%
f_b(y) &=& (y-1)^{\ri(\om-m\Om)}\left\{1-\dfrac{(\om-m\Om)(\om+3m\Om) -	2\ri m\Om - m^2}{4[1 + \ri(\om-m\Om)]}\,(y-1)\right\}. \label{FrobIn2}%
\end{eqnarray}
Then the general solution of the homogeneous Eq.~(\ref{Eq-y}) can be written as:
\be
\label{GenSol}%
f(y) = A_0f_a(y) + B_0f_b(y)
\ee
where $A_0$ and $B_0$ are arbitrary constants. From the physical point of view, $f_a(y)$ describes a co-current (i.e., toward the center) travelling wave which has finite phase speed and wavelength, whereas rapidly oscillating function near the horizon, $f_b(y)$ (due to the pre-factor $(y-1)^{\ri(\om-m\Om)}$), describes a wave which travels against the flow and has a vanishing wavelength at the horizon, because its radial velocity approaches zero in the laboratory reference frame. Bearing this in mind, we impose the boundary condition $B_0=0$, which physically stands for that there is no waves coming from under the horizon, and mathematically this means that $f(y)$ is the analytic function in the neighbourhood of $y=1$.

On the periphery of the flow, where $x \gg 1$, the homogeneous Eq.~(\ref{Eq-x}) can be written in the approximate form:
\[
x^2\dfrac{\dd^2 f}{\dd x^2} + (1 - 2\ri\om)x\,\dfrac{\dd f}{\dd x} +
\Bl[\om^2x^2 - m(m+2\om\Om)\Br]f=0.
\]
Its general solution can be expressed in terms of the Hankel functions (see \cite{Abram}):
\be
f(x)=x^{\ri\om}\Bl[C_1H_\nu^{(1)}(\om x) + C_2H_\nu^{(2)}(\om x)\Br],
\ \ \ \ \nu=(m^2+2m\om\Om-2\om^2)^{1/2},
\label{Sol-x}
\ee
where $C_{1,\,2}$ are constants When $\om x\gg 1$ it has the following asymptotic representation:
\be
f(x)\sim \sqrt{\dfrac{2}{\pi\om}}\,x^{-\frac{1}{2}+\ri\om}
\left\{C_1\re^{\ri[\om x-(2\nu+1)\pi/4]} +
C_2\re^{-\ri[\om x-(2\nu+1)\pi/4]}\right\}.
\label{AsInf}
\ee
Up to a common factor, the constants $C_1$ and $C_2$ can be interpreted as the amplitudes of reflected and incident waves, respectively.

In the conclusion of this section we note that the conservation law (\ref{ConsL}) for the potential disturbances in dimensionless form can be written as:
\be
x\hat{J} = - 2\left(\om - \dfrac{m\Om}{x^2}\right)|f|^2 -
\ri \left(x - \dfrac{1}{x}\right)\left(f^*\dfrac{\dd f}{\dd x} -
f\dfrac{\dd f^*}{\dd x}\right) = {\rm const},
\label{DFlux}
\ee
This equation relates the amplitudes of incident and reflected waves ($C_2$ and $C_1$) with the amplitude $A_0$ of the disturbance at the horizon at $x = 1$. Using Eqs.~(\ref{AsInf}) and (\ref{FrobIn1}), (\ref{FrobIn2}), one can easily find that for $x\gg 1$
\[
x\hat{J} = \dfrac{4}{\pi}\left(\left|C_1\right|^2 - \left|C_2\right|^2\right),
\]
whereas when $x\to 1_{+0}$\ \ $x\hat{J} = -2(\om - m\Om)|A_0|^2$. Hence, the reflection coefficient is:
\be
{\cal R} = \left|\dfrac{C_1}{C_2}\right|^2 = 1 -\dfrac{\pi}{2}
\Bl(\om - m\Om\Br)\left|\dfrac{A_0}{C_2}\right|^2.
\label{Refl}
\ee
Thus, one can see that ${\cal R} > 1$, when Eq.~(\ref{Super}) is satisfied. In this case the over-reflection occurs. Such phenomenon which stems from the ability of a wave to extract energy and momentum from the mean flow was considered for the first time, apparently, by Miles \cite{Miles} and Ribner \cite{Ribner} for acoustic waves as earlier as 1957, and then by many other authors for the different kinds of waves (see, for example, Refs. \cite{Fejer, Jones, Breeding, McKenzie, Lindzen74, Elt-McKenz, Acheson, Dick-Clare, Grisler-Dick}; we cite here only the pioneering works and cannot present the full list of publications on this theme. Some other references can be found in the review \cite{Lindzen88}).

To obtain the dependence of ${\cal R}$ on parameters, we need to find the ratio of coefficients $A_0$ and $C_2$; this will be done analytically in Section \ref{sec:4} and numerically in Section \ref{sec:5}.
\section{AXIALLY SYMMETRIC DISTURBANCES ($m=0$)}
\label{sec:3}
%\hspace*{\parindent}
%
For axially symmetric disturbances Eq.~(\ref{Eq-y}),
\[
(y-1)\dfrac{\dd^2 f}{\dd y^2}+(1 - \ri\om)\dfrac{\dd f}{\dd y}+
\dfrac{\om^2}{4}f = \dfrac{\Pi_0}{2}\left(1-\dfrac{\ri\om}{2}\,y\right)
\re^{-\ri\om y/2},
\]
has the exact analytical solution:
\be
\label{ExactSol}
f(y) = \ri\Pih_0\re^{-\ri\om y/2} + \left(\dfrac{\om}{2}\,
\sqrt{y-1}\right)^{\ri\om}\left[A_1J_{\ri\om}\Bl(\om\sqrt{y-1}\,\Br)
+ A_2J_{-\ri\om}\Bl(\om\sqrt{y-1}\,\Br)\right],
\ee
where $\Pih_0=\Pi_0/\om$, $J_\nu(z)$ is the Bessel function of the first kind \cite{Abram} and $A_1$ and $A_2$ are arbitrary constants. As one can easily see, this solution has two independent components, one of them (the first term) describes the vortex perturbation, whereas the second one describes the potential gravity wave.

Eliminating the rapidly oscillating component when $y\to 1_{+0}$ (i.e., the wave coming from under the horizon), we set $A_1=0$ and obtain (cf. Eq.~(9) in \cite{Dolan11}):
$$
f(x) =  A_2\left(\dfrac{\om}{2}\,\sqrt{y-1}\right)^{\ri\om}
J_{-\ri\om}\Bl(\om\sqrt{y-1}\,\Br) + \ri\Pih_0\re^{-\ri\om y/2}
$$
\be
{}\equiv \dfrac{A_2}{2}\left(\dfrac{\om}{2}\,\sqrt{x^2-1}\right)^{\ri\om}
\left[\re^{-\pi\om}H_{\ri\om}^{(1)}\Bl(\om\sqrt{x^2-1}\Br) +
\re^{\pi\om}H_{\ri\om}^{(2)}\Bl(\om\sqrt{x^2-1}\Br)\right] +
\ri\Pih_0\re^{-\ri\om y/2},
\label{Sol-m0}
\ee
where $H_\nu^{(1,\,2)}(z)$ are the Hankel functions of the first and second kinds \cite{Abram}. Using their asymptotic expansions for $|z|\gg 1$, we see that at the periphery of the flow, for $\om x\gg 1$ (cf. Eq.~(\ref{AsInf})):
$$
f(x) \sim \dfrac{A_2}{\sqrt{4\pi}}\left(\dfrac{\om x}{2}\right)^
{-\frac{1}{2}+\ri\om}\left[\re^{\ri(\om x-\pi/4)-\pi\om/2} +
\re^{-\ri(\om x-\pi/4)+\pi\om/2}\right] + \ri\Pih_0\re^{-\ri\om x^2/2}
$$
\be
{} = x^{-\frac{1}{2}+\ri\om}\left\{C_{out}\re^{\ri\om x} +
C_{in}\re^{-\ri\om x}\right\} + \ri\Pih_0\re^{-\ri\om x^2/2},
\label{Sol0as}
\ee
where $C_{in}$ and $C_{out}$ are the amplitudes of incident and reflected waves.

The solution (\ref{Sol-m0}) is the sum of incident and reflected surface gravity waves, propagating with the dimensionless velocity $\tilde{c}_0=1$ and the PV perturbation (described by the last term), which is simply transported by the flow as a passive scalar impurity without interaction with the gravity waves. Therefore, the reflection coefficient is:
\be
{\cal R}_0 = \left|\dfrac{C_{out}}{C_{in}}\right|^2 = \re^{-2\pi\om} < 1.
\label{Rm0}
\ee
This formula coincides with the earlier derived in \cite{Dolan11, Dolan} for the purely potential perturbations.
\section{DISTURBANCES WITH $m > 0$}
\label{sec:4}
%\hspace*{\parindent}
%
When $m\ne 0$, the scattering problem is much more complicated and can be solved analytically only in the case of low frequencies ($0<\om\ll 1$) which is, however, in the agreement with the shallow-water approximation. Assuming that $0<\Om=O(1)$ and using the method of matched asymptotic expansions (see, for example, \cite{Nayfeh}), we obtain approximate solutions of homogeneous Eqs.~(\ref{Eq-x})--(\ref{Eq-s}) (describing the potential disturbances) in three domains, $x=O(1)$, $\om^{1/2} x=O(1)$, and $\om x =O(1)$, and match them on the boundaries of the domains. Then we construct solutions of non-homogeneous Eqs.~(\ref{Eq-x})--(\ref{Eq-s}), which take into account the PV.
\subsection{Potential disturbances ($\Pih = 0$)}
\label{sec:4.1}
%\hspace*{\parindent}
%
{\bf 1.} In the domain where $x=O(1)$, all terms in the homogeneous equations (\ref{Eq-x})--(\ref{Eq-s}) containing $\om$ are negligibly small. With this in mind, let us put in the homogeneous Eq.~(\ref{Eq-s}) $f(s)=s^{m/2}G(s)$ and introduce the variable $u=1-s$. In the leading order we obtain a hypergeometric equation:
\be
u(1-u)\dfrac{\dd^2 G}{\dd u^2}+\Bl[1+\ri m\Om-(2+m+\ri m\Om)u\Br]
\dfrac{\dd G}{\dd u} - \dfrac{m}{4}(1+\ri\Om)(m+2+\ri m\Om)G=0.
\label{Eqn1}
\ee
The general solution of this equation can be presented in terms of a hypergeometric function $F(a,b;c;z)$ \cite{Abram}:
	\be
	G(u)=A_1\,F(a,1+a;1+\ri m\Om;u)+A_2u^{-\ri m\Om}\,F(a^*,1+a^*;1-\ri m\Om;u),
	\quad a=\dfrac{m}{2}(1+\ri\Om).
	\label{Sol-1f}
	\ee
The first term in this solution is regular in the neighbourhood of the horizon $u=0$, whereas the second term is rapidly oscillating function. Let us start with the regular solution $f_a$. Setting $A_1=1$ and $A_2=0$, we turn back to the variable $s$ and	find that when $x = O(1)$:
	\be
	f_a\approx s^{m/2}F(a,1+a;1+\ri m\Om;1-s) = x^{-m}F(a,1+a;1+\ri m\Om;1-x^{-2}).
	\label{Sol-1}
	\ee

To calculate the asymptotic expansion for $x\gg 1$,	we pass from $1-s$ to $s$ in the argument of function $F$ (see {\cite{Abram}, 15.3.12) and obtain:
$$
f_a(x) \approx \dfrac{\Gm(m)\Gm(1+\ri m\Om)}{\Gm(a)\Gm(1+a)}\,
x^m\sum\limits_{k=0}^{m-1}
\dfrac{(a-m)_k(1+a-m)_k}{(1-m)_k\,k\,!}\,x^{-2k}
$$
$$
{} + \dfrac{(-1)^m\Gm(1+\ri m\Om)}{\Gm(a-m)\Gm(1+a-m)}\,x^{-m}
\sum\limits_{k=0}^{\infty}
\dfrac{(a)_k(1+a)_k}{k\,!(m+k)!}\,x^{-2k}
$$
\be
{} \times \Bl[2\ln x+\psi(k+1)+\psi(m+k+1)-\psi(a+k)-\psi(1+a+k)\Br],
		\label{Sol-1as}
		\ee
where $\Gm(z)$ is the Euler gamma-function, $\psi(z)$ is its logarithmic derivative, and we use the Pochhammer symbols: $(q)_0=1$ and $(q)_k = q(q+1)\dots(q+k-1)$, $k=1,\,2,\,3\dots$ \cite{Abram}. \\
			
{\bf 2.} To obtain the general solution in the domain $\om^{1/2} x=O(1)$, let us use in the homogeneous equation (\ref{Eq-x}) a new variable $t=\om^{1/2} x$, then the equation reads:
$$
\hat{L}_m f\equiv\, t^2\,\dfrac{\dd^2 f}{\dd t^2}+t\,\dfrac{\dd f}{\dd t}
-m^2f = {\cal F}
$$
			\be
			\label{Lm}
{} \equiv \om\,\left[\dfrac{\dd^2 f}{\dd t^2}+\left(2\ri-
			\dfrac{1+2\ri m\Om}{t^2}\right)t\,\dfrac{\dd f}{\dd t}-
			\left(t^2-2m\Om+\dfrac{m^2\Om^2-2\ri m\Om}{t^2}\right)f\right].
			\ee

Consider solution to this equation in the form of a series with respect to small parameter $\om$: $f=f_{m0}+\om f_{m1}+\dots$, then we obtain:
$$
f = B_1t^m+B_2t^{-m}+\om\left\{-\dfrac{B_1}{4}\left[\dfrac{m}{m-1}(1-\ri\Om)(m-2-\ri m\Om)\,t^{m-2}+\dfrac{t^{m+2}}{m+1}
\right]\right.
$$
$$
{} +\dfrac{B_2}{4}\left[\dfrac{m}{m+1}(1+\ri\Om)(m+2+\ri m\Om)\,t^{-m-2} + \dfrac{t^{-m+2}}{m-1}\right]
$$

\be
{} \left. +\,\ri\,\Bl[(1-\ri\Om)B_1t^m+(1+\ri\Om)B_2t^{-m}\Br]\ln\dfrac{t}{\om^{1/2}}\right\} + O(\om^2),
\label{Sol-2}
\ee

where $B_1$ and $B_2$ are arbitrary constants. \\
			
{\bf 3.} In the domain $\om x=O(1)$ Eq.~(\ref{Eq-x}) reduces in the leading
			order to the Bessel equation:
			\[
			\dfrac{\dd^2 f}{\dd x^2} + \dfrac{1}{x}\,\dfrac{\dd f}{\dd x} +
			\left(\om^2 - \dfrac{m^2}{x^2}\right)f = 0.
			\]
The general solution of this equation is:
			\be
			f = C_1J_m(\om x) + C_2Y_m(\om x) =
			C_{in}\,H_m^{(2)}(\om x) + C_{out}\,H_m^{(1)}(\om x),
			\label{Sol-3}
			\ee
where $C_{in}$ and $C_{out}$ are amplitudes of the incident and reflected waves (cf. (\ref{Sol-x})),
			$J_m(z)$, $Y_m(z)$, and $H_m^{(1,2)}(z)$ are the Bessel functions of the
			first, second, and third kind {\cite{Abram}, and $C_1=C_{in}+C_{out}$,\ \
				$C_2 = \ri(C_{out} - C_{in})$. Finally, using the well-known expansions of
				Bessel functions in power series \cite{Abram}, we can rewrite Eq.~(\ref{Sol-3})
				as:
$$
				f = \left(\dfrac{\om x}{2}\right)^m \sum\limits_{k=0}^{\infty}
\dfrac{(-1)^k}{k!(m+k)!}\left(\dfrac{\om x}{2}\right)^{2k}
\left\{C_{in}+C_{out}-\dfrac{\ri}{\pi}(C_{in}-C_{out})\left[
2\ln\dfrac{\om x}{2}-\psi(k+1)\right.\right.
$$
				\be
\left.\left.{\vphantom{\dfrac{\om x}{2}}} -\psi(m+k+1)\right]\vphantom{\dfrac{\om x^W}{2_W}}\right\}
				+\,\dfrac{\ri}{\pi}(C_{in}-C_{out})\left(\dfrac{\om x}{2}\right)^{-m}
				\sum\limits_{k=0}^{m-1}\dfrac{(m-k-1)!}{k!}\left(\dfrac{\om x}{2}\right)^{2k}.
				\label{Sol-3s}
				\ee
\\
				
{\bf 4.} Matching expansions of function $f_a(x)$ (\ref{Sol-1as}) and (\ref{Sol-2}) for $1\ll x\ll\om^{-1/2}$ and taking into account that $a-m=-a^*$ (see Eq.~(\ref{Sol-1f})), we obtain (for details see Appendix~\ref{ap:B}):
$$
B_{1a} \equiv A_{1a}\om^{-m/2}, \quad A_{1a} = \dfrac{(m-1)!\,\Gm(1+\ri m\Om)}{\Gm(a)\,\Gm(1+a)}, \quad 				B_{2a} \equiv A_{2a}\om^{m/2},
$$
				\be
				A_{2a} = \dfrac{(-1)^m\Gm(1+\ri m\Om)}{\Gm(-a^*)\,\Gm(1-a^*)}
				\Bl[\psi(1)+\psi(m+1)-\psi(a)-\psi(1+a)\Br].
				\label{MatchL}
				\ee
				
Matching further expansion (\ref{Sol-2}) with expansion (\ref{Sol-3s}) for $\om^{-1/2}\ll x\ll\om^{-1}$, we obtain:
				\be
\!\!\left\{				
				\ba{l}
				C_{in}^{(a)} = \dfrac{m!}{2}\left(\dfrac{\om}{2}\right)^{-m}\!\!A_{1a}
				+ \dfrac{A_{2a}}{2(m-1)!}\left(\dfrac{\om}{2}\right)^{m}
				\left[2\ln\dfrac{\om}{2}-\psi(1)-\psi(m+1)-\ri\pi\right],
				\\ \\
				C_{out}^{(a)} = \dfrac{m!}{2}\left(\dfrac{\om}{2}\right)^{-m}\!\!A_{1a}
				+ \dfrac{A_{2a}}{2(m-1)!}\left(\dfrac{\om}{2}\right)^{m}
				\left[2\ln\dfrac{\om}{2}-\psi(1)-\psi(m+1)+\ri\pi\right],
				\\ \\
				C_{1a} = C_{in}^{(a)} + C_{out}^{(a)} =
				m!\left(\dfrac{\om}{2}\right)^{-m}\!\!A_{1a} + \dfrac{A_{2a}}{(m-1)!}
				\left(\dfrac{\om}{2}\right)^{m}
				\left[2\ln\dfrac{\om}{2}-\psi(1)-\psi(m+1)\right],
				\\ \\
				C_{2a} = \ri\Bl(C_{out}^{(a)} - C_{in}^{(a)}\Br) =
				- \dfrac{\pi A_{2a}}{(m-1)!}\left(\dfrac{\om}{2}\right)^{m}.
				\label{MatchR}
				\ea
\right.
				\ee
				
Based on these coefficients, we can determine the ratio:
				\be
				R = \dfrac{C_{out}^{(a)}}{C_{in}^{(a)}}=\dfrac{1+\dfrac{A_{2a}/A_{1a}}
					{(m-1)!\,m!}\left(\dfrac{\om}{2}\right)^{2m}\left[2\ln\dfrac{\om}{2}-
					\psi(1)-\psi(m+1)+\ri\pi\right]} {1+\dfrac{A_{2a}/A_{1a}}
					{(m-1)!\,m!}\left(\dfrac{\om}{2}\right)^{2m}\left[2\ln\dfrac{\om}{2}-
					\psi(1)-\psi(m+1)-\ri\pi\right]}.
				\label{R1}
				\ee
				
Using Eq.~(\ref{MatchL}) and taking into account that $\om\ll 1$, we can present this expression as (see Appendix~\ref{ap:B}):
$$
				R \approx 1+\dfrac{\ri}{2\pi}\,\dfrac{|a|^2\,|\Gm(a)|^4}{[(m-1)!]^2\,m!}
\left(\dfrac{\om}{2}\right)^{2m}\left[\re^{\pi m\Om/2}-(-1)^m
\re^{-\pi m\Om/2}\right]^2\times
$$
				\be
\Bl[\psi(1)+\psi(m+1)	-\,\psi(a)-\psi(1+a)\Br].
				\label{R2}
				\ee

Note further that from the well-known representation for $\psi(z)$ \cite{Abram},
				\[
				\psi(z)=-\gamma-\sum\limits_{k=0}^{\infty}\left(\dfrac{1}{z+k}-
				\dfrac{1}{k+1}
				\right),
				\]
				where $\gamma=0.5772156649\dots$ is the Euler constant, it follows that
				\[
				\psi(x+\ri y)-\psi(x-\ri y) = 2\ri y\sum\limits_{k=0}^{\infty}
				\Bl[(x+k)^2+y^2\Br]^{-1}.
				\]
Therefore, the imaginary parts of $\psi(z)$ and $z$ have the same sign, and we finally obtain for the reflection coefficient ${\cal R} \equiv |R|^2$ of potential waves:
				\be
				{\cal R} \approx 1+\dfrac{|a|^2\,|\Gm(a)|^4}{\pi[(m-1)!]^2\,m!}
				\left(\dfrac{\om}{2}\right)^{2m}\left[\re^{\pi m\Om/2}-(-1)^m
				\re^{-\pi
					m\Om/2}\right]^2\IM\Bl[\psi(a)+\psi(1+a)\Br] > 1.
				\label{R3}
				\ee
This expression is greater than one, as expected (see Eq.~(\ref{Refl})). \\
				
{\bf 5.} Now let us set $A_1=0$ and $A_2=1$ in Eq.~(\ref{Sol-1f}) and construct the second solution $f_b$ of homogeneous Eqs.~(\ref{Eq-x})--(\ref{Eq-s}). In the domain where $x=O(1)$, we obtain
$$
				f_b = s^{m/2}(1-s)^{-\ri m\Om}F(a^*,1+a^*;1-\ri m\Om;1-s)+O(\om)
$$
				\be
{} = x^{-m+2\ri m\Om}(x^2-1)^{-\ri m\Om}F(a^*,1+a^*;1-\ri m\Om;1-x^{-2})+O(\om).
				\label{Sol-b1}
				\ee
Passing from $1-s$ to $s$ in the argument of function $F$, we obtain (see Eq. 15.3.12 in \cite{Abram}):
$$
					f_b(x) \approx x^{2\ri m\Om}(x^2-1)^{-\ri m\Om}\left\{
\dfrac{\Gm(m)\Gm(1-\ri m\Om)}{\Gm(a^*)\Gm(1+a^*)}\, x^m\sum\limits_{k=0}^{m-1}
\dfrac{(a^*-m)_k(1+a^*-m)_k}{(1-m)_k\,k\,!}\,x^{-2k}\right.
$$
$$
					+ \dfrac{(-1)^m\Gm(1-\ri m\Om)}{\Gm(-a)\Gm(1-a)}\,x^{-m}\sum\limits_{k=0}^{\infty}
\dfrac{(a^*)_k(1+a^*)_k}{k\,!(m+k)!}\,x^{-2k}
$$
\be
\left.\phantom{\dfrac{Gm}{Gm}}
					\times \Bl[2\ln x+\psi(k+1)+\psi(m+k+1)-\psi(a^*+k)-\psi(1+a^*+k)\Br]\right\}.
					\label{Sol-2as}
					\ee
					
Matching Eq.~(\ref{Sol-2as}) with Eq.~(\ref{Sol-2}) for $1\ll x\ll\om^{-1/2}$ leads to the equalities:
					\be
					A_{1b} = A^*_{1a}, \qquad A_{2b} = A^*_{2a}.
					\label{Sol-b2}
					\ee

And matching Eq.~(\ref{Sol-2as}) with Eq.~(\ref{Sol-3s}) for $\om^{-1/2}\ll x\ll\om^{-1}$
					gives the amplitude factors in front of Bessel functions. These factors can be
alternatively obtained by substitution of Eqs.~(\ref{Sol-b2}) into Eqs.~(\ref{MatchR}):
$$
					f_b \approx 2^{m-1}m!A^*_{1a}\om^{-m}\left\{\left[1+\dfrac{A^*_{2a}/A^*_{1a}}
{m!(m-1)!}\left(\dfrac{\om}{2}\right)^{2m}\left(2\ln\dfrac{\om}{2}-\psi(1)-
\psi(m+1)+\ri\pi\right)\right]H_m^{(1)}(\om x)\right.
$$					\be
\left.
					+\left[1+\dfrac{A^*_{2a}/A^*_{1a}}
					{m!(m-1)!}\left(\dfrac{\om}{2}\right)^{2m}\left(2\ln\dfrac{\om}{2}-\psi(1)-
					\psi(m+1)-\ri\pi\right)\right]H_m^{(2)}(\om x)\right\}.
					\label{Sol-b3}
					\ee
					
Thus,  functions $f_a(x)$ and $f_b(x)$ are described by the approximate formulae
					\be
					f_a(x) = \left\{
					\ba{ll}
					x^{-m}F(a,1+a;1+\ri m\Om;1-x^{-2}), & x=O(1), \\
					A_{1a}x^m+A_{2a}x^{-m}, & \om^{1/2}x=O(1), \\
					C_{in}^{(a)}\,H_m^{(2)}(\om x) + C_{out}^{(a)}\,H_m^{(1)}(\om x),
					& \om x=O(1),
					\ea
					\right.
					\label{fa}
					\ee
					\be
					f_b(x) = \left\{
					\ba{ll}
					x^{-m+2\ri m\Om}(x^2-1)^{-\ri m\Om}
					F(a^*,1+a^*;1-\ri m\Om;1-x^{-2}), & x=O(1), \\
					A_{1a}^*x^m+A_{2a}^*x^{-m}, & \om^{1/2}x=O(1), \\
					C_{in}^{(b)}\,H_m^{(2)}(\om x) + C_{out}^{(b)}\,H_m^{(1)}(\om x),
					& \om x=O(1),
					\ea
					\right.
					\label{fb}
					\ee
where $A_{1a}$ and $A_{2a}$ are determined by Eqs.~(\ref{MatchL}),
					$C_{in}^{(a)}$ and $C_{out}^{(a)}$ are determined by Eqs.~(\ref{MatchR}),
					and $C_{in}^{(b)}$ and $C_{out}^{(b)}$ are obtained from the last two by
					replacing $A_{1,2a}$ with $A^*_{1,2a}$.
					
Having two linear independent solutions $f_a(x)$ and $f_b(x)$, we can calculate the Wronskian of these functions, $W[f_a(x), f_b(x)] \equiv f'_af_b -f_af'_b$. The Wronskian will be needed in the next subsection, when we will construct a particular solution of non-homogeneous equations (\ref{Eq-x})--(\ref{Eq-s}). Taking into account the relationship between the variables $x$, $y$, and $s$ (see Eqs.~(\ref{Eq-y}), (\ref{Eq-s})), one can easily see that
					\[
					W_x = 2xW_y = -2x^{-3}W_s,
					\]
where the subscript index indicates on which variable the derivative is taken. The Wronskian of Eq.~(\ref{Eq-s}) has the form
					\[
					W_s = W_{0s}s^{-(1+\ri\om)}(1-s)^{\ri(\om-m\Om)-1}, \quad
					W_{0s}={\rm const}.
					\]
In Luke's notations (see Eq. 6.6.(18) in Luke \cite{Luke}), $w_1(u)=F(a,1+a;1+\ri m\Om;u)$ and
					$w_2(u)=u^{\ri m\Om}F(a^*,1+a^*;1+\ri m\Om;u)$. The Wronskian of these functions with respect to the variable $u$ is:
					\[
					w_1w'_2 - w'_1w_2 = -W_0u^{-\ri m\Om-1}(1-u)^{-(m+1)}=
					-W_0s^{-(m+1)}(1-s)^{-\ri m\Om-1}, \quad W_0=\ri m\Om.
					\]
					Since $u=1-s$,\ \ $\dd/\dd s=-\dd/\dd u$,\ \ $W_{0s}=-W_0$, therefore finally we have:
					\be
					W_x = 2W_0x^{1+2\ri m\Om}(x^2-1)^{\ri(\om-m\Om)-1}.
					\label{Wx}
					\ee
\subsection{Disturbances carrying a potential vorticity ($\Pih \ne 0$)}
\label{sec:4.2}
%\hspace*{\parindent}
					%
Now let us calculate the contribution of non-zero potential vorticity into the process of wave scattering. In Section~\ref{sec:3}, we have shown that for axially-symmetric disturbances ($m=0$) this contribution has no effect on the dynamics of gravity waves, but its magnitude, $\Pih_0=\Pi_0/\om$, does not decrease with the distance and therefore at the far periphery of the flow, when $\om x\gg 1$, gravity waves become invisible on its background (see
					Eq.~(\ref{Sol0as})). It is easy to verify that when $m\ne 0$, the PV
					contribution does not decrease with the distance as well. Indeed, Eq.~(\ref{Eq-x})
					with the right-hand side (\ref{Rxys}) has a particular solution $f_{PV}(x)$, which for $\om x^2\gg 1$ has the asymptotic expansion:
					\[
					f_{PV}(x) = \ri\Pih_0x^{\ri m\Om}\re^{-\ri\om x^2/2}\left[1+m\left(
					\dfrac{\Om}{\om x^2}+\dfrac{2\ri\Om-m(1-\Om^2)}{\om^2x^4}+
					O(\om^{-3}x^{-6})\right)\right].
					\]
Note that function $f_{PV}(x)$ does not satisfy the boundary condition at the horizon $x = 1$ and therefore
does not contain gravity waves.

					Let us extract a non-decreasing with radius part of the disturbance.
					Putting
					\be
					f(x)=\ri\Pih_0\Bl(m\,f_1(x) + x^{\ri m\Om}\re^{-\ri\om x^2/2}\Br)
					\label{fPi}
					\ee
					and substituting it into Eq.~(\ref{Eq-x}), we obtain the equation
					\be
					\hat{L}_x f_1=\Bl[m(1+\Om^2)-\om\Om x^2\Br]x^{\ri m\Om+2}
					\re^{-\ri\om x^2/2}.
					\label{Eq-f1}
					\ee
					
Let us construct a solution to this equation containing neither waves emitted from the domain $x<1$ (i.e., coming from under the horizon), no incident waves coming from the periphery, that is a solution describing {\it emission of gravity waves stimulated by the PV}.
					
					The desired solution can be written in the form
					\be
					f_1(x) = \left(D-\dfrac{I_b(x)}{2W_0}\right)f_a(x)-
					\dfrac{I_a(x)}{2W_0}\,f_b(x),
					\label{f1}
					\ee
					where
					\be
					\left\{\ba{l}
					I_a(x) = \din_{1}^{x}\!\xi^{-\ri m\Om-1}
					(\xi^2-1)^{-\ri(\om-m\Om)}\Bl[m(1+\Om^2)-\om\Om\xi^2\Br]
					f_a(\xi)\re^{-\ri\om\xi^2/2}\dd\xi\,,
					\\ \\
					I_b(x) = \II{x}\!\xi^{-\ri m\Om-1}
					(\xi^2-1)^{-\ri(\om-m\Om)}\Bl[m(1+\Om^2)-\om\Om\xi^2\Br]
					f_b(\xi)\re^{-\ri\om\xi^2/2}\dd\xi\,.
					\ea\right.
					\label{Iab}
					\ee
The parameter $D$ in Eq.~(\ref{f1}) should be chosen in such a way that the incident wave has the zero amplitude,
$$
D = \dfrac{I_a(\infty)\,C_{in}^{(b)}}{2W_0\,C_{in}^{(a)}}
$$
					\be
					\approx
					-\dfrac{\ri I_a(\infty)A_{1a}^*}{2m\Om A_{1a}}\left[1+\dfrac{(\om/2)^{2m}}
					{(m-1)!\,m!}\left(\dfrac{A_{2a}^*}{A_{1a}^*}-\dfrac{A_{2a}}{A_{1a}}\right)
					\left(2\ln\dfrac{\om}{2}-\psi(1)-\psi(m+1)-\ri\pi\right)\right].
					\label{D}
					\ee
					
In accordance with the structure of functions $f_{a,\,b}$, we divide the integration domain for $I_a(x)$ into three parts,\ \ $x=O(1)$,\ \ $\om^{1/2}x=O(1)$, and $\om x=O(1)$.\\
					
					{\bf 1.} When $x=O(1)$, we set in the leading order $\om=0$ and passing to the variable $s=1/x^2$, we obtain:
					\[
					I_{a1} \approx \dfrac{m}{2}(1+\Om^2)\IE{s}\!u^{a^*-1}
					(1-u)^{\ri m \Om}F(a,1+a;1+\ri m\Om;1-u)\,\dd u\,.
					\]
Introducing the new variable of integration, $t=1-u$ and using the formula (see Eq. 1.15.3.8. in \cite{Prud3})
					\[
					\int t^{c-1}(1-t)^{b-c-1}F(a,b;c;t)\,\dd t\, = c^{-1}t^c(1-t)^{b-c} F(a+1,b;c+1;t),
					\]
					we find
					\be
					I_{a1} \approx \dfrac{m(1+\Om^2)}{2(1+\ri m\Om)}\,s^{a^*}(1-s)^{1+\ri m\Om}
					F(1+a,1+a;2+\ri m\Om;1-s).
					\label{a1}
					\ee
					
On passing from $1-s$ to $s$, we see that the expansion of $I_{a1}$ in terms of $s=x^{-2}$ (see Appendix~\ref{ap:B}) does not contain a constant term then, keeping only the most rapidly increasing with $x$ term, we obtain:
					\be
					I_{a1} \sim \dfrac{(1+\Om^2)m\,!\Gm(1+\ri m\Om)}{2[\Gm(1+a)]^2}\,x^{2a}.
					\label{Ia1}
					\ee\\
					
{\bf 2.} To find the contribution of the domain where $\om^{1/2}x=O(1)$ to $I_a$, it is necessary to calculate the integral
					\[
					I(x)=\int x^{-\ri m\Om-1}(x^2-1)^{-\ri(\om-m\Om)}\Bl[m(1+\Om^2)-
					\om\Om x^2\Br]f_a(x)\re^{-\ri\om x^2/2}\,\dd x\,.
					\]
					On passing to $y=x^2$ and bearing in mind that $x\gg 1$, we obtain
					\[
					I \approx \dfrac{1}{2}\din y^{\ri m\Om/2-1}\Bl[m(1+\Om^2)-
					\om\Om y\Br]\Bl(A_{1a}y^{m/2}+A_{2a}y^{-m/2}\Br)\re^{-\ri\om y/2}\,\dd y\,.
					\]

Due to the presence of a rapidly oscillating exponential, the contribution
					of the $\om x=O(1)$ domain can also be taken into account by extending the
					upper limit of integration to $x=+\infty$. On passing to $t=\ri\om y/2$,
					we see that the term with $A_{2a}$ is negligible and the integral reduces to
					\[
					I(x)\approx\dfrac{1}{2}\left(\dfrac{\ri\om}{2}\right)^{-a}\left\{D_a-A_{1a}
					\II{\ri\om x^2/2}t^{a-1}\re^{-t}\Bl[m(1+\Om^2)+2\,\ri\,\Om\,t\Br]\,\dd t
					\right\},
					\]
					where $D_a$ is determined by the equality $I(x)=I_a(x)$. The integral can
					be expressed in terms of a confluent hypergeometric function (see Eq. 6.9(21) in \cite{HTF1}),
						\[
						I \approx \dfrac{1}{2}\left(\dfrac{\ri\om}{2}\right)^{-a}\!\left\{D_a\!-
						\!A_{1a}\re^{-\ri\om x^2/2}\left[m(1\!+\!\Om^2)
						\Psi\left(1\!-\!a,1\!-\!a;\dfrac{\ri\om x^2}{2}\right)\!+\!
						2\,\ri\,\Om\,\Psi\left(-a,-a;\dfrac{\ri\om x^2}{2}\right)\right]
						\right\}.
						\]
						
						Further, we pass from  function $\Psi(a,c;z)$ to the Kummer functions $\Phi(a,c;z)$
						and use the following equation (see Eq. 6.5(7) in \cite{HTF1}):
							\[
							\Psi(a,c;z)=\dfrac{\Gm(1-c)}{\Gm(a-c+1)}\Phi(a,c;z)+
							\dfrac{\Gm(c-1)}{\Gm(a)}z^{1-c}\Phi(a-c+1,2-c;z)
							\]
							and the well-known formula $\Phi(a,a;z)=\re^z$ to obtain:
							\[
							\ba{l}
							I(x) \approx \left(\dfrac{D_a}{2} - A_{1a}\Gm(1+a)\right)
							\left(\dfrac{\ri\om}{2}\right)^{-a}
							\\ \\ \phantom{I}%\left.\left.
							+\,\dfrac{A_{1a}}{2}\,x^{2a}\re^{-\ri\om x^2/2}\left[\dfrac{m(1+\Om^2)}{a}
							\,\Phi\left(1,1+a;\dfrac{\ri\om x^2}{2}\right)-\dfrac{\om\Om x^2}{1+a}
							\,\Phi\left(1,2+a;\dfrac{\ri\om x^2}{2}\right)\right].
							\ea
							\]
							
							Calculating the asymptotic of $I(x)$ for $\om x^2\ll 1$,
							\[
							I(x)\sim\left(\dfrac{D_a}{2} - A_{1a}\Gm(1+a)\right)
							\left(\dfrac{\ri\om}{2}\right)^{-a}+\dfrac{A_{1a}}{2a}\,m(1+\Om^2)x^{2a}
							\Bl[1+O(\om x^2)\Br],
							\]
							comparing it with Eq.~(\ref{Ia1}), and using Eq.~(\ref{MatchL}), we find
							\be
							\dfrac{D_a}{2} = A_{1a}\Gm(1+a), \qquad
							I_a(\infty)=\dfrac{D_a}{2}\left(\dfrac{\ri\om}{2}\right)^{-a}=
							\dfrac{(m-1)!\,\Gm(1+\ri m\Om)}{\Gm(a)}
							\left(\dfrac{\ri\om}{2}\right)^{-a}.
							\label{Ia}
							\ee
							
							Now we use Eqs.~(\ref{D}), (\ref{f1}), (\ref{fa}), and (\ref{fb}) to get the
							asymptotic representation of solution in the wave zone $\om x=O(1)$:
							\[
							f_1(x) \approx Df_a(x)-\dfrac{I_a(\infty)}{2W_0}\,f_b(x)\sim
							CH_m^{(1))}(\om x), \quad
							C\approx \dfrac{\pi I_a(\infty)}{2\Om m!}\left(\dfrac{\om}{2}\right)^m
							A_{1a}^*\left(\dfrac{A_{2a}}{A_{1a}}-\dfrac{A^*_{2a}}{A^*_{1a}}\right).
							\]
							Then, using Eq.~(\ref{Ia}), expression for the ratio $A_{2a}/A_{1a}$ (see
							Appendix~\ref{ap:B}), and the known formula
							$|\Gm(1+\ri m\Om)|^2=\pi m\Om/\sinh(\pi m\Om)$, we find the
							amplitude of emitted wave normalized by $\Pih_0$ (see (\ref{fPi}))
							\be
							% \ba{l}
							A=\ri mC\approx \dfrac{m\ri^{-m/2}}{2}\,\dfrac{\re^{\pi m\Om/2}-
								(-1)^m\re^{-\pi m\Om/2}} {\re^{\pi m\Om/2}+(-1)^m\re^{-\pi m\Om/2}}
							\left(\dfrac{\om}{2}\right)^{a^*}\!\!\re^{\pi m\Om/4}\Gm(1+a)
							%    \\ \\ \phantom{uu}\times\,
							\IM\Bl[\psi(a)+\psi(1+a)\Br]\,.
							% \ea
							\label{f1as}
							\ee
							
							The asymptotic of the complete solution of Eq.~(\ref{Eq-x}) describing
							PV-stimulated emission of gravity waves now is:
							\be
							f_\Pi(x) \sim \Pih_0\left\{\ri x^{\ri m\Om}\re^{-\ri\om x^2/2} +
							A\,H_m^{(1)}(\om x)\right\}.
							\label{fPas}
							\ee
							
							To obtain the solution in the entire domain $x>1$, it is necessary to find
							the integral $I_b(x)$ (see Eq.~(\ref{Iab})). Calculations similar to those made
							above for two different domains yield: \\
a) for the domain $\om x^2=O(1)$:
							\be
							\ba{l}
							I_{b}(x) \approx
							-\dfrac{A^*_{1a}}{2}\left[\dfrac{m(1+\Om^2)}{a}
							\,\Phi\left(1,1\!+\!a;\dfrac{\ri\om x^2}{2}\right)-\dfrac{\om\Om x^2}{1+a}
							\,\Phi\left(1,2\!+\!a;\dfrac{\ri\om x^2}{2}\right)\right]
							x^{2a}\re^{-\ri\om x^2/2}
							\\ \\ %\phantom{Iw}%
							+\,A^*_{1a}\left(\dfrac{\ri\om}{2}\right)^{-a}\!\!\Gm(1+a)
							\,\,\stackrel{\om x^2\ll 1}{\sim}\,\,
							A^*_{1a}\left(\dfrac{\ri\om}{2}\right)^{-a}\!\!\Gm(1+a)
							-\dfrac{m(1+\Om^2)}{2a}A^*_{1a}x^{2a}\Bl[1+O(\om x^2)\Br];
							\ea
							\label{Ib}
							\ee
b) and for the domain $x=O(1)$:
							\[
							\ba{l}
							I_{b1}(x) \approx D_b+\dfrac{\ri m^2\Om(1+\Om^2)}{2|a|^2}x^{-2a^*}
							F(a^*,a^*;-\ri m\Om;1-x^{-2})
							\\ \\ \phantom{www}
							\stackrel{x\gg 1}{\sim}\,
							D_b-\dfrac{m!(1+\Om^2)\Gm(1-\ri m\Om)}{2|a|^2[\Gm(a^*)]^2}
							\,x^{2a}+O(x^{2a-2}).
							\ea
							\]
							
Comparing this with Eq.~(\ref{Ib}), we find that
							\[
							D_b=A^*_{1a}\left(\dfrac{\ri\om}{2}\right)^{-a}\!\Gm(1+a)
							=\dfrac{(m-1)!\,\Gm(1+a)\Gm(1-\ri m\Om)}{\Gm(a^*)\Gm(1+a^*)}
							\left(\dfrac{\ri\om}{2}\right)^{-a}.
							\]
							At the horizon $x=1$, with the relationship $m^2(1+\Om^2)=4|a|^2$ taken into account,
							we obtain
							\be
							I_b(1) \approx \dfrac{(m-1)!\,\Gm(1+a)\Gm(1-\ri m\Om)}{\Gm(a^*)\Gm(1+a^*)}
							\left(\dfrac{\ri\om}{2}\right)^{-a}+2\ri\Om,
							\label{Ib1}
							\ee
							where the second term is negligible compared to the first one.
							
In the conclusion to this section we note that when $m = 0$, solution of non-homogeneous Eq. (\ref{ExactSol}) consists, as usual, of particular solution (PS) of non-homogeneous equation representing a potential vorticity term and general solution (GS) of the homogeneous equation describing surface waves. Herewith, the PS does not contain potential surface waves at the infinity (when $r \to \infty$) and satisfies boundary condition the horizon, i.e. remains regular when $r \to 1_+$. 

When $m > 0$ one can construct again a PS which does not contain surface waves at the infinity, but such solution is non-physical as it is singular at the horizon. In contrast to that solution (\ref{f1}) satisfies the boundary condition at the horizon and contains the potential vorticity through the integral terms $I_a$ and $I_b$ as per Eqs. (\ref{Iab}). The $D$-term in Eq. (\ref{f1}) represents the amplitude of the solution to the homogeneous equation. The PS regular at the horizon in this case has the asymptotic (\ref{fPas}) when $r \to \infty$. This asymptotic solution consists of a vortex component and a potential component representing a surface wave radiated outside of vortex region with the amplitude proportional to $m$ (see Eq. (\ref{f1as})).

							\section{NUMERICAL RESULTS}
							\label{sec:5}
							%\hspace*{\parindent}
							%
 The analytical solutions obtained above describe both the over-reflection and PV-stimulated emission of gravity waves. However, these solutions are valid only in the limit of small frequencies, when formally $\om/\Om\ll 1$. They demonstrate, in particular, the effect of over-reflection per se (see Eq.~(\ref{R3})), but do not provide information about a frequency range where the over-reflection occurs. To learn more about the dynamics of perturbations containing both potential and vortex components in a wide range of frequencies, it is necessary to solve Eq.~(\ref{Eq-f1}) numerically for arbitrary $\om/\Om$. To this end it is convenient to substitute in the equation $f_1(x) = x^{\ri m\Om}\re^{-\ri\om/2}G(x)$ and change the variable $\xi=\sqrt{x^2-1}$; then we obtain:
							\be
							\dfrac{\dd^2G}{\dd\xi^2}+\Bl[1-2\ri(\om-m\Om)\Br]\dfrac{1}{\xi}
							\dfrac{\dd G}{\dd\xi}+\left[\om^2-\dfrac{m^2(1+\Om^2)}{1+\xi^2}\right]G =
							\left[\dfrac{m(1+\Om^2)}{1+\xi^2}-\om\Om\right]\re^{-\ri\om\xi^2/2}.
							\label{Eq-g}
							\ee
Function in the right-hand side does not have a singularity on the horizon, hence, in agreement with Section~\ref{sec:2}, we exclude the emission from under the horizon, if we require the analyticity of function $G(\xi)$ in the neighbourhood of $\xi=0$.

Setting $G(0)=1$, we first find a solution of the homogeneous Eq.~(\ref{Eq-g}) with the zero right-hand side. Such solution describes scattering of purely potential disturbances (i.e., gravity waves). Figure~\ref{f02}a) shows the dependence of the reflection coefficient ${\cal R}$ on the frequency for $m = 1$ and various values of angular velocity $\Om$, whereas in Fig.~\ref{f02}b) the same dependence is shown for $\Om = 2$ and various values of $m$.%\\%
%	Fig. 2
\begin{figure}[h!]
\centerline{\includegraphics[width=15cm]{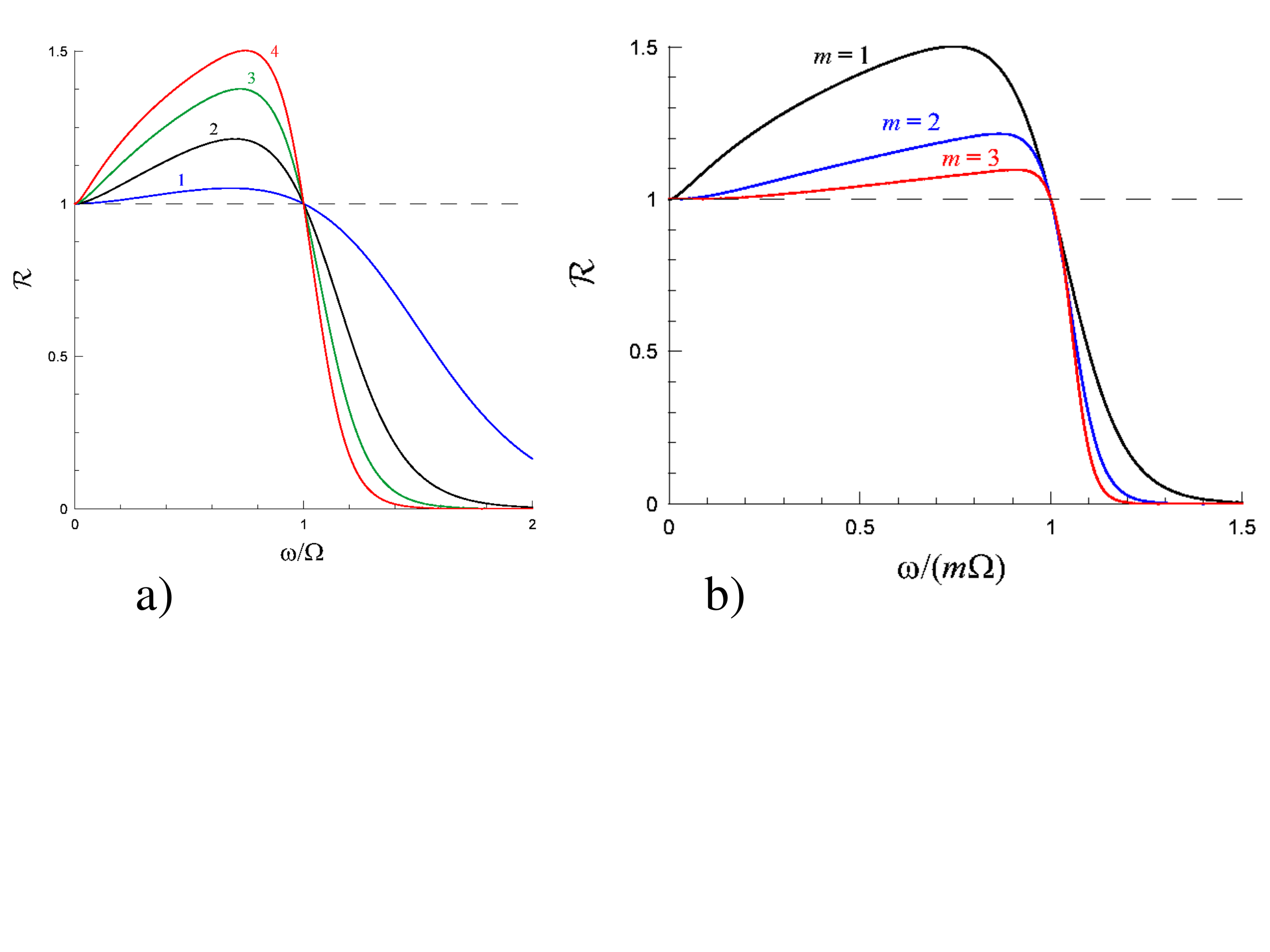}} %
\vspace*{-4.0cm}%
\caption{\protect\footnotesize
a) Reflection coefficient of gravity waves with the azimuthal number $m = 1$ as the function of normalised frequency $\om/\Om$ for $\Om=0.5$ (line 1),\ \ $\Om=1$ (line 2),\ \ $\Om=1.5$ (line 3), and\ \ $\Om=2$ (line 4). b) Reflection coefficient of gravity waves as the function of normalised frequency $\om/(m\Om)$ for $\Om = 2$ and few values of azimuthal number $m$.}
\label{f02}
\end{figure}

%	Fig. 3
\begin{figure}[h!tb]
\centerline{\includegraphics[width=15cm]{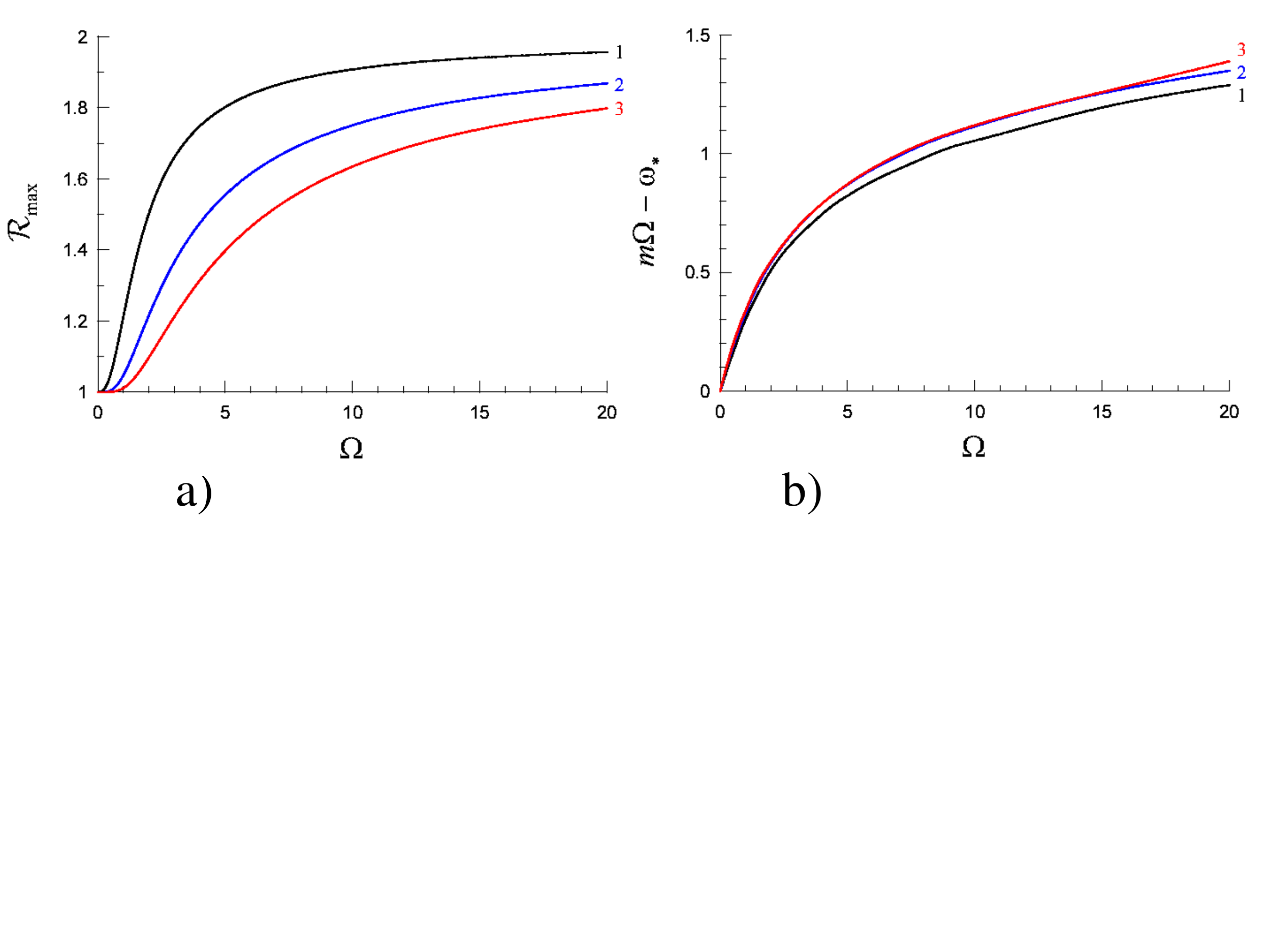}} %
\vspace*{-5.0cm}%
\caption{\protect\footnotesize
Dependences of maximal reflection coefficient (a) and corresponding wave frequency $\om_*$ (b) on $\Om$ for few several values of azimuthal number $m = 1,\; 2,\; 3$.}
\label{f03}
\end{figure}
							
As it would be expected, ${\cal R}>1$ when Eq.~(\ref{Super}) is satisfied, and it turns to unity in the cases $\om\to 0$ (in the complete agreement with Eq.~(\ref{R3})) and $\om\to m\Om$. For each $\Om$ and $m$, the reflection coefficient attains a maximal value ${\cal R}_{\rm max}(m,\Om)$ at a certain frequency $\om=\om_*(m,\Om)<m\Om$. In Fig.~\ref{f03} the dependences of ${\cal R}_{\rm max}$ and $(m\Om-\om_*)$ on $\Om$ are shown for various values of $m$. It is seen that whilst ${\cal R}_{\rm max}$ grows with $\Om$, the larger is $m$, the smaller is the growth rate. In any case, ${\cal R}_{\rm max}$ asymptotically approaches 2 regardless of $m$ when $\Om \to \infty$.
							
							At the next step we solve the complete (non-homogeneous) Eq.~(\ref{Eq-g})
							with the boundary condition $G(0)=1 $ and then, find such a value of $G(0)$
							for which the solution of this equation does not contain an incident
							gravity wave. Figure~\ref{f04} shows the dependence of normalized amplitude $|A|/\Pih_0$ of emitted wave on frequency. The amplitude goes to zero when $\om \to 0$ and when $\om \to \infty$; it attains a maximum at $\om=O(m\Om)$. The maximum increases with $\Om$ the faster, the larger $m$ is, and does not vanish even when $\Om\to 0$ -- see Fig.~\ref{f05}. In Fig.~\ref{f04}\,b one can see that for $\om\ll m\Om$\ \ $|A|\sim\om^{m/2}$ in agreement with Eq.~(\ref{f1as}).
%	Fig. 4
\begin{figure}[h!tb]
\centerline{\includegraphics[width=15cm]{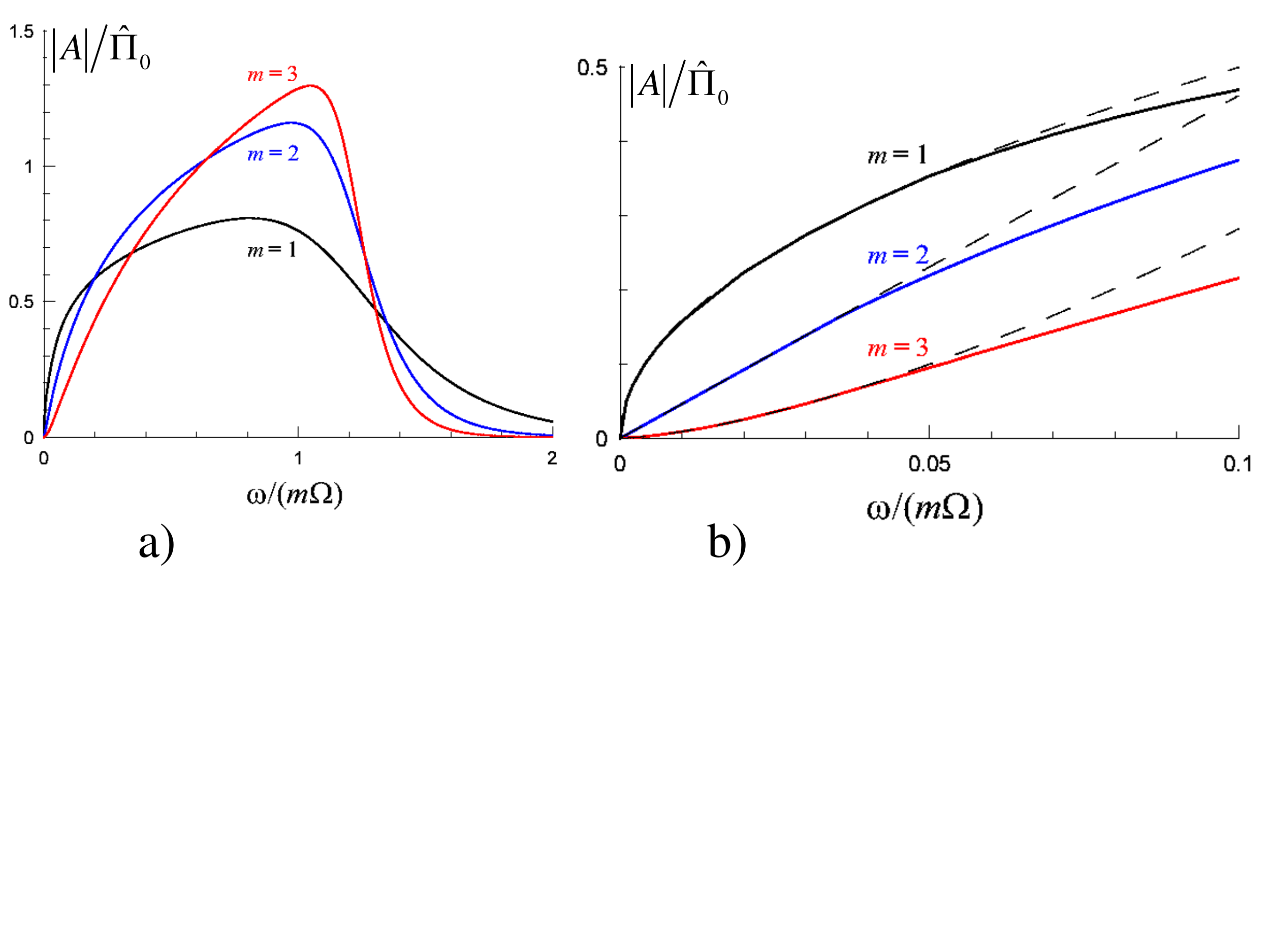}} %
\vspace*{-5.0cm}%
\caption{\protect\footnotesize Dependence of normalized amplitude of PV-stimulated gravity wave on the normalised frequency for $\Om=1$ and various numbers $m$. Frame (a) shows a whole dependence, and frame (b) shows a fragment for $\om\ll m\Om$. Dashed lines in frame (b) show the analytical dependences $|A|\sim\om^{m/2}$ for $\om/(m\Om) \ll 1$. }
\label{f04}
\end{figure}
%	Fig. 5
\begin{figure}[h!tb]
\centerline{\includegraphics[width=15cm]{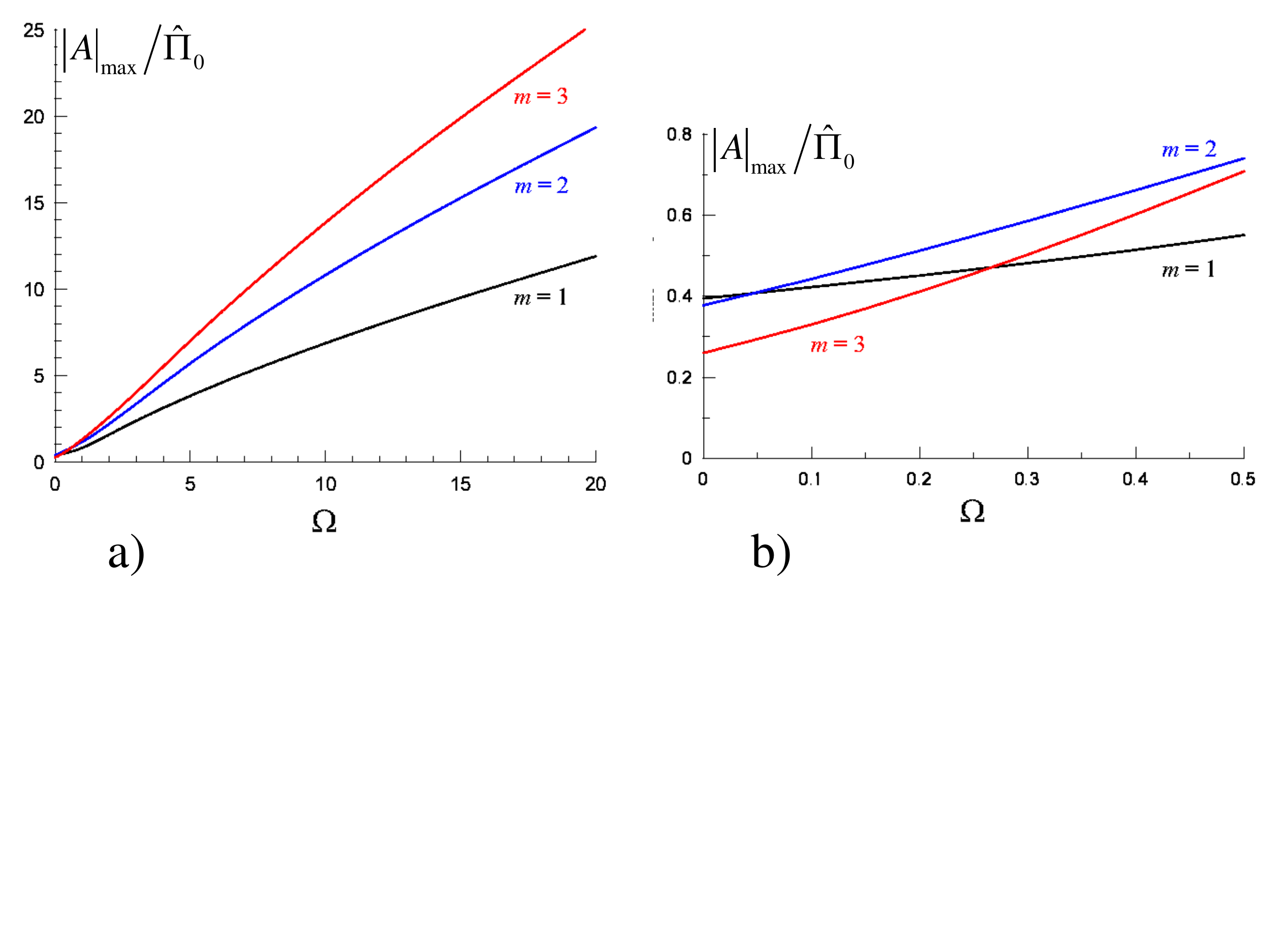}} %
\vspace*{-5.0cm}%
\caption{\protect\footnotesize Maximal amplitude of PV-stimulated gravity wave as the function of $\Om$
for various values of $m$. Frame (a) shows this dependence in the wide range of $\Om\le 20$, and frame (b) shows a fragment for $\Om\le 0.5$.}
\label{f05}
\end{figure}
\section{DISCUSSION}
\label{sec:6}

This paper is devoted to investigation of some aspects of wave dynamics in a DBT flow, which was originally proposed in Ref.~\cite{SchUn} and currently is widely used as an analogue of a rotating black hole. Using the method of matched asymptotic expansions, we have calculated analytically the reflection coefficient ${\cal R}$ for low-frequency purely potential wave disturbances (surface gravity waves) scattered by a DBT vortex and have demonstrated that over-reflection takes place (${\cal R}>1$, see Eq.~(\ref{R3})), when the condition $\om < m\Omega$ is fulfilled. It should be noted that preceding efforts to calculate the reflection coefficient for the low frequencies and $m \ne 0$ in the Born approximation \cite{Dolan11, Dolan} actually yielded ${\cal R} = 1$ or, at most, slightly greater than one, so that the over-reflection of gravity waves was calculated only numerically \cite{Dolan, Wein15} and observed experimentally \cite{Torres17}. 

The results obtained in our paper can be compared with the experimental results published in Ref. \cite{Torres17}. As has been shown in that paper (see also \cite{StepYeoh}), the angular velocity profile of the bathtub vortex can be well-approximated by the Lamb vortex (known also as the Burgers and Rankine vortex \cite{Lamb, Lautrup}):
\be
V_\varphi = \Omega_0\frac{r_0^2}{r}\left[1 - \exp{\left(-\frac{r_0^2}{r^2}\right)}\right],
\label{Burgers}
\ee
where $\Omega_0 = 69.4$ rad/s and $r_0 = 1.34$ cm.

Incoming waves were generated at the frequency range from $f = 2.87$ Hz to $f = 4.11$ Hz so that $0.13 \le \omega/\Omega_0 \le 0.37$, where $\omega = 2\pi f$. For this interval our {\it inviscid} theory predicts ${\cal R}_1 \approx 1.1 - 1.35$ for $m = 1$, and ${\cal R}_2 \approx 1 - 1.05$ for $m = 2$ (see Fig. \ref{f02}). In the experiment \cite{Torres17}, where the viscosity effect was quite noticeable, it was obtained ${\cal R}_1 \approx 1.09 \pm 0.03$ and ${\cal R}_2 \approx 1.14 \pm 0.08$ (see Fig. 2 in \cite{Torres17}). Such agreement between the theory and experiment can be accepted as satisfactory, taking into account influence of viscosity in the laboratory set-up (the authors promise to reduce the viscosity effect in the future experiments).

 As has been mentioned in the Introduction, the effective curved space-time associated with a potential DBT flow provides an incomplete analogue for the Kerr metric describing a real rotating black hole. In Ref.~\cite{VW05} it was shown in detail that in order to have a closer (but still incomplete!) analogue, one should consider a flow with a vorticity. However, the vorticity leads to the linkage of various eigenmotions of the flow, and the wave equation (\ref{dAlamb}) acquires the right-hand side due to vorticity. In the meantime, the vorticity is affected by surface gravity waves (see Refs.~\cite{Visser18, Visser01}), and we have a coupled wave-vorticity self-consistent dynamics.

 As the first step towards studying such an interplay, disturbances, containing both gravity waves and potential vorticity, were considered on the DBT flow. Because the basic flow is potential one, PV disturbances are not affected by waves. They are transported by the flow in accordance with Eq.~(\ref{Pi}) and simultaneously emit gravity waves as described by non-homogeneous equation (\ref{PerEq}). Using solutions obtained for purely potential waves (i.e., within the homogeneous equation (\ref{PerEq})), we have constructed the analytical solution describing PV stimulated emission of low-frequency ($\om\ll m\Om$) gravity waves and calculated the emissivity (see Eq.~(\ref{f1as})). Then we have extended this analysis by means of numerical calculations to a wider frequency range $\om\lesssim m\Om$ (see Figs.~\ref{f04} and \ref{f05}).
 
 In the conclusion we draw the reader attention to the specific feature of observation of a wave scattering by a DBT vortex. In general, any perturbation contains both gravity waves and PV disturbances, therefore the over-reflection and PV-stimulated emission of gravity waves occur simultaneously. As the result of these combined effects, the ``reflection coefficient'' ${\cal R}_{obs}$ measurable at the periphery of the flow (the ratio of squared amplitudes of outgoing and incident gravity waves) can take any value, in principle, from zero to infinity. It is clear that the noticeable difference between ${\cal R}_{obs}$ and the genuine reflection coefficient of {\it strictly potential} disturbances ${\cal R}$ requires a proper level of PV.  To this end it should be a way to control the level of PV. An appropriate way may be based on the fact that at the periphery of the flow, PV disturbances look like small-scale ripples, which, in contrast to gravity waves, do not decrease with the radius (see Eqs.~(\ref{Sol0as}) and (\ref{fPas})). However, there are at least two circumstances making difficult control of the PV level. Firstly, all measurements are usually conducted at a finite (and not very large) distances from the flow centre, whereas the gravity wave magnitude decreases with the radius not so fast, only as $r^{-1/2}$. Secondly, Fig.~\ref{f05} demonstrates that the effect of stimulated emission is the higher, the greater the angular velocity $\Om$ of the flow at the horizon and the azimuthal number $m$ are. For this reason, ${\cal R}_{obs}$ can significantly differ from ${\cal R}$ even at a relatively low level of PV.
 
 When this work has been completed and published in Phys. Rev. Fluids (2019, v. 4, n. 3, 034704) we became aware about a very relevant paper \cite{Berti} where the reflection coefficient of acoustic waves from the draining bathtub fluid flow was calculated numerically. Results obtained in our paper are in a good agreement with the results published by the authors of Ref. \cite{Berti}. Many other important features of a classical model of a black hole were discussed in that paper. 
\begin{acknowledgments}
S.C. was supported by Budgetary Funding of Basic Research Program of the State Academies of Russia, Project No. II.16. Y.S. acknowledges the funding of this study from the State task program in the sphere of scientific activity of the Ministry of Education and Science of the Russian Federation (Project No. 5.1246.2017/4.6) and grant of the President of the Russian Federation for state support of leading scientific schools of the Russian Federation (NSH-2685.2018.5).
\end{acknowledgments}

\appendix
 \section{Conservation laws for perturbations}
\label{ap:A}
\hspace*{\parindent}
Equations (\ref{PerEq}) and (\ref{Pi}) can be derived from the principle of least action:
\be
\ba{l}
\delta S = 0, \quad S = \din{\cal L}\left(
\dfrac{\partial\upsilon}{\partial t}\,,\,\nabla\upsilon,\,\dfrac{\partial\Pi}{\partial t}\,,\,\nabla\Pi,\,\lambda;\,r\right)\dd t\,\dd^2 x,
\ea
\label{Lagr}
\ee
where the Lagrangian density is:
\be
\ba{l}
{\cal L} = |\hat{L}\upsilon|^2 - gH_0\left(\left|
\dfrac{\partial\upsilon}{\partial r}\right|^2 +
\left|\dfrac{\partial\upsilon}{r\partial\vp}\right|^2\right) +
g r H_0^2\left(\Pi^*\dfrac{\partial\upsilon}{\partial r}+
\Pi\,\dfrac{\partial\upsilon^*}{\partial r}\right)
+ \lambda^*\hat{L}\Pi + \lambda\hat{L}\Pi^*.
\ea
\label{LagrDen}
\ee
Here the operator $\hat{L} = \dfrac{\partial}{\partial t} + U(r)\dfrac{\partial}{\partial r} +
\dfrac{C}{r^2}\,\dfrac{\partial}{\partial\vp}$ is the same as after Eq. (\ref{Pi}), and $\lambda$ is the Lagrange multiplier obeying the equation
\be
\dfrac{\partial\lambda}{\partial t}+\dfrac{1}{r}\,\dfrac{\partial}{\partial r}
\left(rU\lambda\right)+\dfrac{C}{r^2}\,\dfrac{\partial\lambda}{\partial\vp}-
grH_0^2\dfrac{\partial\upsilon}{\partial r} = 0.
\label{LagrMult}
\ee

Since the action $S$ is invariant under the transformation $(\upsilon,\,\Pi,\,\lambda)\to(\upsilon,\,\Pi,\,\lambda)\,\re^{\ri\al}$ and shift in time, $t\to t+\tau$, where $\al$ and $\tau$ are real constants, then in accordance with the Noether theorem (see, for example, \cite{Sardan}), the wave action conserves:
\be
\dfrac{\partial N}{\partial t} + \dfrac{1}{r}\dfrac{\partial}{\partial r}
\left(rJ_{Nr}\right) + \dfrac{1}{r}\dfrac{\partial J_{N\vp}}{\partial\vp} = 0,
\label{WAc}
\ee
\begin{eqnarray}
N &=& \ri\left[\upsilon^*\hat{L}\upsilon - \upsilon\hat{L}\upsilon^* +
\lambda\Pi^*-\lambda^*\Pi\right], \nonumber\\
%&& \nonumber \\
J_{Nr} &=& NU - \ri gH_0\left(\upsilon^*\dfrac{\partial\upsilon}{\partial r} -
\upsilon\,\dfrac{\partial\upsilon^*}{\partial r}\right) -
\ri grH_0^2(\Pi^*\upsilon - \Pi\upsilon^*), \nonumber\\
%&& \nonumber \\
J_{N\vp} &=& \dfrac{C}{r}\,N - \dfrac{\ri gH_0}{r}\left(
\upsilon^*\dfrac{\partial\upsilon}{\partial\vp} -
\upsilon\,\dfrac{\partial\upsilon^*}{\partial\vp}\right). \nonumber
\end{eqnarray}

The wave energy (more precisely, pseudo-energy, see \cite{McInt}) conserves too:
\be
%\ba{l}
\dfrac{\partial{\cal E}}{\partial t} + \dfrac{1}{r}\dfrac{\partial\left(rJ_{Er}\right)}{\partial r}
+ \dfrac{1}{r}\dfrac{\partial J_{E\vp}}{\partial\vp} = 0,
\label{WEn}
\ee
\begin{eqnarray}
{\cal E} &=& \left|\dfrac{\partial\upsilon}{\partial t}\right|^2 - \left|U\dfrac{\partial\upsilon}{\partial r} + \dfrac{C}{r^2}\,
\dfrac{\partial\upsilon}{\partial\vp}\right|^2 + gH_0\left(\left|\dfrac{\partial\upsilon}{\partial r}
\right|^2 + \dfrac{1}{r^2}\left|\dfrac{\partial\upsilon}{\partial\vp}\right|^2\right) -
rgH_0^2\left(\Pi^*\dfrac{\partial\upsilon}{\partial r} + \Pi\dfrac{\partial\upsilon^*}{\partial r}\right) \nonumber\\
{} &-& \,\lambda^*\left(U\dfrac{\partial\Pi}{\partial r} + \dfrac{C}{r^2}\,
\dfrac{\partial\Pi}{\partial\vp}\right) - \lambda\left(U\dfrac{\partial\Pi^*}
{\partial r} + \dfrac{C}{r^2}\,\dfrac{\partial\Pi^*}{\partial\vp}\right), \nonumber\\
%&& \nonumber \\
J_{Er} &=& 2U\left|\dfrac{\partial\upsilon}{\partial t}\right|^2 +
\dfrac{CU}{r^2}\left(\dfrac{\partial\upsilon}{\partial t} \,
\dfrac{\partial\upsilon^*}{\partial\vp} +
\dfrac{\partial\upsilon^*}{\partial t} \,\dfrac{\partial\upsilon}{\partial\vp}\right) -
\Bl(gH_0 - U^2\Br)\left(\dfrac{\partial\upsilon}{\partial t} \,
\dfrac{\partial\upsilon^*}{\partial r}
+ \dfrac{\partial\upsilon^*}{\partial t} \,\dfrac{\partial\upsilon}{\partial r}\right) \nonumber\\
{} &+&\,U\left(\lambda^*\dfrac{\partial\Pi}{\partial t}+\lambda\dfrac{\partial\Pi^*}
{\partial t}\right) +rgH_0^2\left(\Pi^*\dfrac{\partial\upsilon}{\partial t}+
\Pi\dfrac{\partial\upsilon^*}{\partial t}\right), \nonumber\\
%&& \nonumber \\
J_{E\vp} &=& \dfrac{2C}{r}\left|\dfrac{\partial\upsilon}{\partial t}\right|^2 +
\dfrac{CU}{r}\left(\dfrac{\partial\upsilon}{\partial t} \,
\dfrac{\partial\upsilon^*}{\partial r} +
\dfrac{\partial\upsilon^*}{\partial t} \,\dfrac{\partial\upsilon}{\partial r}\right) -
\dfrac{1}{r}\left(gH_0 - \dfrac{C^2}{r^2}\right)\left(
\dfrac{\partial\upsilon}{\partial t} \,\dfrac{\partial\upsilon^*}{\partial\vp} +
\dfrac{\partial\upsilon^*}{\partial t} \,\dfrac{\partial\upsilon}{\partial\vp}\right) \nonumber\\
{} &+& \,\dfrac{C}{r}\left(\lambda^*\dfrac{\partial\Pi}{\partial t} + \lambda\dfrac{\partial\Pi^*}{\partial t}\right), \nonumber
\end{eqnarray}
where the asterisk denotes complex conjugation.

In the case of a monochromatic perturbation, $(\upsilon,\,\Pi,\,\lambda) \sim
\exp[\ri(m\vp-\omh t)]$,
the quantities $N$, $J_{Nr}$, $J_{N\vp}$, ${\cal E}$, $J_{Er}$, and $J_{E\vp}$
are independent of $t$ and $\vp$, and the conservation laws (\ref{WAc}) and
(\ref{WEn}) are reduced to the equations for corresponding radial fluxes,
\begin{eqnarray}
rJ_{Nr} &=& 2\left(\omh - \dfrac{mC}{r^2}\right)rU(r)|\upsilon|^2 -
\ri\, r\Bl[gH_0 - U^2(r)\Br]\left(\upsilon^*\dfrac{\dd \upsilon}{\dd r} -
\upsilon\dfrac{\dd \upsilon^*}{\dd r}\right) \nonumber\\
{} &-& \,\ri gr^2H_0^2\Bl(\Pi^*\upsilon-\Pi\upsilon^*\Br) - \ri Ur\Bl(\lambda^*\Pi-\lambda\Pi^*\Br) = {\rm const}, \label{Flux1} \\
rJ_{Er} &\equiv& \omh rJ_{Nr} = {\rm const}. \label{Flux2}
\end{eqnarray}%
In particular, for the potential disturbances ($\Pi=0$) we obtain Eq.~(\ref{ConsL}).							
\section{MATCHING AND AUXILIARY CALCULATIONS}
\label{ap:B}
\subsection{Matching the solutions}
%
%\hspace*{\parindent}
%
When matching the solution in the intermediate domain (Eq.~(\ref{Sol-2})) with the solutions on the left (Eq.~(\ref{Sol-1as})) and on the right
							(Eq.~(\ref{Sol-3s})), we need to equate the coefficients of the terms containing
							$x^m$ and $x^{-m}$ only. In this case the terms in the finite sums in
							Eqs.~(\ref{Sol-1as}) and (\ref{Sol-3s}), as well as the terms with
							exponents between $(-m)$ and $m$ and logarithmic terms are assumed to be
							matched automatically. We verify this by the example of matching the
							solutions described by Eqs.~(\ref{Sol-1as}) and (\ref{Sol-2}).
							
First of all, let us take into account that equation $\hat{L}_m f=t^n$ has the solution
								\[
								f(t) = \dfrac{t^n}{n^2-m^2}\,\ \ \mbox{when} \ \ n^2 \ne m^2\quad \mbox{and} \quad
								f(t) = \pm\dfrac{t^{\pm m}}{2m}\,\ln\dfrac{t}{\om^{1/2}},\ \ \mbox{when} \ \
								n=\pm m.
								\]
Substitution of $f(t)=At^n\equiv\om^{n/2}Ax^n$\ \ ($A={\rm const}$) into the right-hand side of Eq.~(\ref{Lm}) yields
								\[
								\ba{l}
								{\cal F} = \om A\Bl\{\Bl[(n-1-\ri m\Om)^2-1\Br]t^{n-2} +
								2\ri(n+\ri m\Om)t^n - t^{n+2}\Br\}
								\\ \\ \phantom{w}
								= \om^{\frac{n}{2}}\Bl[(n-1-\ri m\Om)^2-1\Br]Ax^{n-2} +
								2\ri\om^{\frac{n}{2}+1}(n+\ri m\Om)Ax^n - \om^{\frac{n}{2}+2}Ax^{n+2}.
								\ea
								\]
Note that, in terms of the variable $x$, the first term in ${\cal F}$ has the same order in $\om$ as $f$, and the next two terms are of a higher order of smallness.

Let us construct now a solution of Eq.~(\ref{Lm}) by the method of successive iterations, setting $f_{m0}=B_{m1}t^m\equiv\om^{m/2}B_{m1}x^m$ and keeping in each iteration only the leading terms. In the $k$-th iteration the leading term has, evidently, the form $A_kx^{m-2k}$, where $A_0=\om^{m/2}B_{m1} = O(1)$, and all $A_k$ are, obviously, of the order of unity too. Then after some algebra we obtain
								\[
								A_{k+1} = \dfrac{(a-m+k)(a-m+k+1)}{(1-m+k)(k+1)}\,A_k,\ \ 0\le k\le m-2.
								\]
Taking into account Eq.~(\ref{MatchL}), one can see that these are the coefficients of finite sum in Eq.~(\ref{Sol-1as}). In addition to that, at $k=m$ we obtain the term containing $A_m x^{-m}\ln x$, where
								\[
								A_m = -2\dfrac{a(a-1)}{m}\,A_{m-1} =
								\dfrac{(-1)^m\Gm(a)\Gm(a+1)}{\Gm(a-m)\Gm(a-m+1)}\,\om^{m/2}B_{m1}.
								\]
This corresponds to the leading logarithmic term in Eq.~(\ref{Sol-1as}). The other terms in Eq.~(\ref{Sol-1as}) can be obtained in a similar way.
					
And the last remark in conclusion to this subsection. The right-hand side of Eq.~(\ref{Sol-2}) contains the similar term proportional to $B_{m2}t^{-m}\ln(t/\om^{1/2})$, but, by virtue of Eq.~(\ref{MatchL}), it is of the higher order of smallness in $\om$ in terms of the variable $x$.
							\subsection{Transition from Eqs.~(\ref{R1}) to (\ref{R2})}
							%
							%\hspace*{\parindent}
							%
							As follows from Eqs.~(\ref{MatchL}),
							\[
							\ba{l}
							\dfrac{A_{2a}}{A_{1a}} = \dfrac{(-1)^m\Gm(a)\Gm(1+a)}{(m-1)!\,\Gm(-a^*)
								\Gm(1-a^*)}\,\Bl[\psi(1)+\psi(m+1)-\psi(a)-\psi(1+a)\Br]
							\\ \\ \phantom{wwi}
							= \dfrac{(-1)^m\Gm(a)\Gm(1+a)\Gm(1+a^*)\Gm(a^*)}{(m-1)!\,\Gm(1+a^*)\Gm(-a^*)
								\Gm(1-a^*)\Gm(a^*)}\,\Bl[\psi(1)+\psi(m+1)-\psi(a)-\psi(1+a)\Br]
							\\ \\ \phantom{wwi}
							= -\dfrac{(-1)^m|a|^2|\Gm(a)|^4}{\pi^2(m-1)!}\,\sin^2(\pi a^*)
							\,\Bl[\psi(1)+\psi(m+1)-\psi(a)-\psi(1+a)\Br].
							\ea
							\]
							Further, we observe that
							\[
							\ba{l}
							\sin(\pi a^*) = \dfrac{1}{2\ri}\Bl[\re^{\ri\pi m(1-\ri\Om)/2} -
							\re^{-\ri\pi m(1-\ri\Om)/2}\Br] = \dfrac{1}{2\ri}\,\re^{\ri\pi m/2}
							\Bl[\re^{\pi m\Om/2} - (-1)^m \re^{-\pi m\Om/2}\Br],
							\\ \\
							\sin^2(\pi a^*) = -\dfrac{(-1)^m}{4}\Bl[\re^{\pi m\Om/2} -
							(-1)^m \re^{-\pi m\Om/2}\Br]^2,
							\ea
							\]
							and obtain Eq.~(\ref{R2}).
							\subsection{Asymptotic expansion of $I_{a1}$}
							%
							%\hspace*{\parindent}
							%
							Passing in Eq.~(\ref{a1}) from $(1-s)$ to $s$ yields (see Eq. 15.3.12 in \cite{Abram})
							\[
							\ba{l}
							I_{a1} \approx \frac{1}{2}(1\!+\Om^2)\Gm(1+\ri m\Om)(1-s)^{1+\ri m\Om}
							\left\{\dfrac{m!\,s^{-a}}{[\Gm(1+a)]^2}\sum\limits_{k=0}^{m-1}
							\dfrac{[(1+a-m)_k]^2}{(1-m)_k\,k!}\,s^k \right.
							\\ \\ \phantom{ww} \left.
							- \,\dfrac{(-1)^m m\,s^{a^*}}{[\Gm(1-a^*)]^2}\,\sum\limits_{k=0}^{\infty}
							\dfrac{[(1\!+\!a)_k]^2}{k!(m\!+\!k)!}\,s^k \Bl[\ln s\!-\!\psi(k\!+\!1)\!-\!
							\psi(m\!+\!k\!+\!1)\!+\!2\psi(1\!+\!a\!+\!k)\Br]\right\}.
							\ea
							\]
							Then, on passing to the variable $x$, we obtain
							\[
							\ba{l}
							I_{a1} \approx \frac{1}{2}(1\!+\Om^2)\Gm(1\!+\!\ri m\Om)x^{-2(1+\ri m\Om)}
							(x^2\!-\!1)^{1+\ri m\Om}
							\left\{\dfrac{m!\,x^{2a}}{[\Gm(1+a)]^2}\sum\limits_{k=0}^{m-1}
							\dfrac{[(1+a-m)_k]^2}{(1-m)_k\,k!}\,x^{-2k} \right.
							\\ \\ \phantom{ww} \left.
							+ \,\dfrac{(-1)^m m\,x^{-2a^*}}{[\Gm(1-a^*)]^2}\,\sum\limits_{k=0}^{\infty}
							\dfrac{[(1\!+\!a)_k]^2}{k!(m\!+\!k)!}\,x^{-2k}\Bl[2\ln x\!+\!\psi(k\!+\!1)\!+\!
							\psi(m\!+\!k\!+\!1)\!-\!2\psi(1\!+\!a\!+\!k)\Br]\right\}
							\ea
							\]
							and finally Eq.~(\ref{Ia1}).


\begin{thebibliography}{99.}
	% 1
	\bibitem{Batchelor}
G.~K.~Batchelor, {\it An Introduction to Fluid Dynamics} (Cambridge University Press, 1967--2002).
	% 2
\bibitem{Dolotin}
V.~V.~Dolotin and A.~M.~Fridman, Generation of an observable turbulence spectrum and solitary dipole vortices in rotating gravitating systems, Zh. Eksp. Teor. Fiz. {\bf 99}, 3--21 (1991) (Engl. transl.: Sov. Phys. JETP, {\bf 72}(1), 1--10, 1991).
	% 3
\bibitem{StepFabr}
Yu.~A.~Stepanyants and A.~L.~Fabrikant, {\it Propagation of Waves in Hydrodynamic Shear Flows} (Moscow, Nauka-Fizmatlit, 1996, in Russian) (English version: A.~L.~Fabrikant and Yu.~A.~Stepanyants, {\it Propagation of Waves in Shear Flows} (World Scientific, Singapore, 1998).
	%  4
\bibitem{StepYeoh}
Y. A. Stepanyants and G. H. Yeoh, Stationary bathtub vortices and a critical regime of liquid discharge, J. Fluid Mech. {\bf 604}, 77–98 (2008).
	% 5
\bibitem{Torres17}
T.~Torres, S.~Patrick, A.~Coutant, M.~Richartz, E.~W.~Tedford, and S.~Weinfurtner, Rotational superradiant scattering in a vortex flow, Nature Phys. {\bf13}, 833--836 (2017).
	%  6
\bibitem{ArtBH}
{\it Artifical black holes}, edited by M. Novello, M. Visser, and G. Volovik (World Scientific, Singapore, 2002).
	%  7
\bibitem{AnalogGrav}
C. Barcel\'{o}, S. Liberati, and M. Visser, Analogue gravity, Living Rev. Relativity {\bf 14}, 3--159 (2011).
	%  8
\bibitem{Unruh81}
W. G. Unruh, Experimental black-hole evaporation? Phys. Rev. Lett. {\bf 46}, 1351--1353 (1981).
	%  9
	\bibitem{Coutant-14}
	A. Coutant and R. Parentani, Undulations from amplified low frequency surface waves, Phys. Fluids {\bf 26}, 044106
	(2014).
	%  10
	\bibitem{Robertson-16}
	S. Robertson, F. Michel, and R. Parentani, Scattering of gravity waves in subcritical flows over an obstacle,
	Phys. Rev. D {\bf 93}, 124060 (2016).
	%  11
	\bibitem{Coutant-16}
	A. Coutant and S. Weinfurtner, The imprint of the analogue Hawking effect in subcritical flows, Phys. Rev. D {\bf 94},
	064026 (2016).
	%  12
	\bibitem{Philbin}
	T. G. Philbin, An exact solution for the Hawking effect in a dispersive fluid, Phys. Rev. D {\bf 94}, 064053 (2016).
	%  13
	\bibitem{ChuErStep}
	S. Churilov, A. Ermakov, and Y. Stepanyants, Wave scattering in spatially inhomogeneous flows, Phys. Rev. D {\bf 96}, 064016 (2017).
	%  14
\bibitem{NJP08}
G.~Rousseaux,  C.~Mathis, P.~Ma{\"{\i}}ssa, T.~G.~Philbin, and U.~Leonhardt, Observation of negative-frequency waves in a water
tank: a classical analogue to the Hawking effect? New J.~Phys. {\bf 10}, 053015 (2008).
	%  15
\bibitem{NJP10}
G.~Rousseaux, P.~Ma{\"{\i}}ssa, C.~Mathis,  P.~Coullet, T.~G.~Philbin, and U.~Leonhardt, Horizon effects with surface waves
on moving water, New J.~Phys. {\bf 12}, 095018 (2010).
	%  16
\bibitem{Weinfurtner}
S. Weinfurtner, E. W. Tedford, M. C. J. Penrice, W. G. Unruh, and G. A. Lawrence, Measurement of stimulated Hawking emission in an analogue system, Phys. Rev. Lett. {\bf 106}, 021302 (2011).
	%  17
\bibitem{Euve}
L.-P. Euv\'e, F. Michel, R. Parentani, T. G. Philbin, and G. Rousseaux, Observation of noise
correlated by the Hawking effect in a water tank, Phys. Rev. Lett. {\bf 117}, 121301 (2016).
	%  18
\bibitem{Steinhauer}
J. Steinhauer, Observation of quantum Hawking radiation and its entanglement in an analogue black hole, Nat. Phys. {\bf 12}, 959 (2016).
	%  19
	\bibitem{Chandra}
	S. Chandrasekhar, {\it The Mathematical Theory of Black Holes} (Oxford University Press, New York, 1983).
	% 20
	\bibitem{PressT}
	W. H. Press and S. A. Teukolsky, Floating orbits, superradiant
	scattering and the black-hole bomb, Nature {\bf 238}, 211--212 (1972).
	% 21
	\bibitem{Star}
	A. A. Starobinsky, Amplification of waves during reflection from a
	rotating ``black hole", Sov. Phys. JETP {\bf 37}, 28--32 (1973).
	% 22
	\bibitem{StarCh}
	A. A. Starobinsky and S. M. Churilov, Amplification of electromagnetic and gravitational waves scattered by a rotating ``black hole", Sov. Phys. JETP {\bf 38}, 1--5 (1974).
	% 23
\bibitem{KopLeon}
V.~F.~Kop'ev and E.~A.~Leont'ev, On acoustic instability of axial vortex, Akust. Zhurnal {\bf 29}, n.~2, 192--198 (1983).
(Engl. transl.: Sov. Phys. Acoust., {\bf 29}(2), 111--115, 1983).
	% 24
\bibitem{BroadMoore}
E.~G.~Broadbent and D.~W.~Moore, Acoustic destabilization of vortices,  Phil. Trans. Roy. Soc. London {\bf A 290}, 
n. 1372, 353--371 (1979).  
	% 25
\bibitem{GolFabr}
G.~M.~Golemshtok and A.~L.~Fabrikant, Scattering and amplification of sound waves by cylindrical vortex, Akust. Zhurnal {\bf 26}, n.~3, 383--390 (1980). (Engl. transl.: Sov. Phys. Acoust., {\bf 26}(3), 209--213, 1980).
	%  26
	\bibitem{Brito15}
	R. Brito, V. Cardoso, and P. Pani, Superradiance, Lect. Notes Phys. {\bf 906}, 1--237 (2015).
	%  27
	\bibitem{SchUn}
	R. Sch\"{u}tzhold and W.G. Unruh, Gravity wave analogs of black holes, Phys. Rev. D {\bf 66}, 044019 (2002).
	%  28
	\bibitem{VW05}
M. Visser and S. Weinfurtner, Vortex analogue for the equatorial geometry of the Kerr black hole, Class. Quant. Grav. {\bf 22}, 2493--2510 (2005).
	%  29
	\bibitem{Dolan11}
    S. R. Dolan, E. S. Oliveira, and L. C. B. Crispino, Aharonov--Bohm effect in a draining bathtub vortex, Phys. Lett. B {\bf 701}, 485--489 (2011).
	% 30
	\bibitem{Dolan}
	S. R. Dolan and E. S. Oliveira, Scattering by a draining bathtub vortex, Phys. Rev. D {\bf 87}, 124038 (2013).
	%  31
	\bibitem{Wein15}
	M. Richartz, A. Prain, S. Liberati, and S. Weinfurtner, Rotating black holes in a draining bathtube: superradiant scattering of gravity	waves, Phys. Rev. D {\bf 91}, 124018 (2015).
	% 32
	\bibitem{Visser18}
	S. Liberati, S. Schuster, G. Tricella, and M. Visser, Vorticity in analogue spacetimes, arXiv:1802.04785v2 [gr-qc] (2018).
	%  33
	\bibitem{Visser01}
	S. E. Perez Bergliaffa, K. Hibberd, M. Stone, and M. Visser, Wave equation for sound in
	fluids with vorticity, Physica D {\bf 191}, 121--136 (2004).
	% 34
	\bibitem{Dolzh}
	F. V. Dolzhansky, {\it Fundamentals of Geophysical Hydrodynamics} (Springer, Heidelberg, 2013).
	% 35
\bibitem{Goldstein}
M. E. Goldstein, {\it Aeroacoustics} (McGrawHill, New York, 1976).
	% 36
	\bibitem{Ince}
	E. L. Ince, {\it Ordinary Differential Equations} (Dover Publications, 1958).
	% 37
\bibitem{Abram}
{\it Handbook of Mathematical Functions}, edited by M. Abramowitz and I. A. Stegun (National Bureau of Standards, 1964).
	% 38
	\bibitem{Miles}
	J. W. Miles, On the reflection of sound at an interface of relative motion, J. Acoust. Soc. Am. {\bf 29}, 226--228 (1957).
	% 39
	\bibitem{Ribner}
	H. S. Ribner, Reflection, transmission and amplification of sound by a moving medium, J. Acoust. Soc. Am. {\bf 29}, 435--441 (1957).
	% 40
\bibitem{Fejer}
J. A. Fejer, Hydromagnetic reflection and refraction at a fluid velocity discontinuity, Phys. Fluids {\bf 6}, 508--512 (1963).
	% 41
\bibitem{Jones}
W. L. Jones, Reflexion and stability of waves in stably stratified fluids with shear flow: a numerical study, J. Fluid Mech. {\bf 34}, 609--624.(1968).
	% 42
\bibitem{Breeding}
R. J. Breeding, A non-linear investigation of critical levels for internal atmospheric gravity waves, J. Fluid Mech. {\bf 50}, 545--563 (1971).
	% 43
\bibitem{Dick-Clare}
R. E. Dickinson, F. J. Clare, Numerical study of the unstable modes of a hyperbolic-tangent barotropic shear flow, J. Atmos. Sci. {\bf 30}, 1035--1049 (1973).
	% 44
\bibitem{McKenzie}
J. F. McKenzie, Reflection and amplification of acoustic-gravity waves at a density and velocity discontinuity, J. Geophys. Res. {\bf 77}, 2915--2926 (1972).
	% 45
\bibitem{Grisler-Dick}
J. E. Geisler, R. E. Dickinson, Numerical study of an interacting Rossby wave and barotropic zonal flow near a critical level, J. Atmos. Sci. {\bf 31}, 946--955 (1974).
	% 46
\bibitem{Lindzen74}
R. S. Lindzen, Stability of a Helmholtz profile in a continuously stratified, infinite Boussinesq fluid -- applications to clear air turbulence, J. Atmos. Sci. {\bf 31}, 1507--1514 (1974).
	% 47
\bibitem{Elt-McKenz}
I. A. Eltayeb, J. F. McKenzie, Critical-level behaviour and wave amplification of a gravity wave incident upon a shear layer, J. Fluid Mech. {\bf 72}, 661--671 (1975).
	% 48
\bibitem{Acheson}
D. J. Acheson, On over-reflexion, J. Fluid Mech. {\bf 77}, 433--472 (1976).
	% 49
\bibitem{Lindzen88}
R. S. Lindzen, Instability of plane parallel shear flow (Toward a mechanistic picture of how it works), PAGEOPH {\bf 126}, n. 1, 103--121 (1988).
	% 50
	\bibitem{Nayfeh}
	A. H. Nayfeh, {\it Introduction in Perturbation Techniques}	(Wiley-Interscience, New York, 1981).
	% 51
	\bibitem{Luke}
	Y. L. Luke, {\it Mathematical Functions and their Approximations} (Academic Press, New York, 1975).
	% 52
	\bibitem{Prud3}
	A. P. Prudnikov, Yu. A. Brychkov, and O. I. Marichev {\it Integrals and Series}, Vol. 3 (Gordon and Breach, New York, 1990).
	% 53
	\bibitem{HTF1}
	A. Erd\'{e}lyi, F. Oberhettinger, and F. G. Tricomi, {\it Higher Transcendental Functions}, Vol. I (McGraw-Hill, New York, 1953).
	%54
\bibitem{Lamb}
H. Lamb, {\it Hydrodynamics}, 6th edn. (Cambridge University Press, 1932).
	%55
\bibitem{Lautrup}
B. Lautrup, {\it Physics of Continuous Matter: Exotic and Everyday Phenomena in the Macroscopic
	World}, (IoP Publishing, 2005).
	% 56
	\bibitem{Sardan}
	G. Sardanashvily, {\it Noether's theorems. Applications in Mechanics and Field Theory} (Atlantis Press, New York, 2016).
	% 57
	\bibitem{McInt}
	M. E. McIntyre, On the `wave momentum' myth, {\it J.~Fluid Mech.}, {\bf	106}, 331 -- 347, 1981.
	%58
	\bibitem{Berti}
E. Berti, V. Cardoso, and J.P.S Lemos, Quasinormal modes and classical wave propagation in analogue black holes, {\it Phys. Rev. D}, {\bf 70}, n. 12, 124006, 2004.
%
\end{thebibliography}
\end{document}